\documentclass[journal]{IEEEtran}

\pdfoutput=1

\usepackage{amssymb}
\usepackage{latexsym}
\usepackage{url}
\usepackage{xcolor}
\usepackage{graphicx}
\usepackage{booktabs}
\usepackage{cite}
\usepackage{bbding}
\usepackage{amsmath}
\usepackage{amsfonts}
\usepackage{cases}
\usepackage{algorithmic}
\usepackage{algorithm}
\usepackage{array}
\usepackage{multirow}
\usepackage{threeparttable}
\usepackage[colorlinks,linkcolor=black]{hyperref}
\usepackage[version=4]{mhchem}
\usepackage{amsmath,nccmath}

\begin{document}

\title{Image Reconstruction with Low-rankness and Self-consistency of \textit{k}-space Data in Parallel MRI}%

\author{Xinlin~Zhang,
        Di~Guo,
        Yiman~Huang,
        Ying~Chen,
        Liansheng~Wang,
        Feng~Huang,
        Xiaobo~Qu*
\thanks{This work was supported in part by National Key R\&D Program of China (2017YFC0108703), National Natural Science Foundation of China (61571380, 61871341, 61811530021, U1632274, 61672335 and 61671399), Natural Science Foundation of Fujian Province of China (2018J06018), Fundamental Research Funds for the Central Universities (20720180056), Science and Technology Program of Xiamen (3502Z20183053), and China Scholarship Council (201806315010). \emph{*Corresponding author: Xiaobo Qu, email address: quxiaobo@xmu.edu.cn.}}%
\thanks{Xinlin Zhang, Yiman Huang, Ying Chen, Xiaobo Qu are with Department of Electronic Science, Fujian Provincial Key Laboratory of Plasma and Magnetic Resonance, School of Electronic Science and Engineering, National Model Microelectronics College, Xiamen University, Xiamen 361005, China.}%
\thanks{Di Guo is with School of Computer and Information Engineering, Fujian Provincial University Key Laboratory of Internet of Things Application Technology, Xiamen University of Technology, Xiamen 361024, China.}%
\thanks{Liansheng Wang is with Department of Computer Science, School of Information Science and Engineering, Xiamen University, Xiamen 361005, China.}%
\thanks{Feng Huang is with Neusoft Medical System, Shanghai 200241, China.}}%

\maketitle

\begin{abstract}
Parallel magnetic resonance imaging has served as an effective and widely adopted technique for accelerating scans. The advent of sparse sampling offers aggressive acceleration, allowing flexible sampling and better reconstruction. Nevertheless, faithfully reconstructing the image from limited data still poses a challenging task. Recent low-rank reconstruction methods exhibit superiority in providing a high-quality image. However, none of them employ the routinely acquired calibration data for improving image quality in parallel magnetic resonance imaging. In this work, an image reconstruction approach named STDLR-SPIRiT was proposed to explore the simultaneous two-directional low-rankness (STDLR) in the k-space data and to mine the data correlation from multiple receiver coils with the iterative self-consistent parallel imaging reconstruction (SPIRiT). The reconstruction problem was then solved with a singular value decomposition-free numerical algorithm. Experimental results of phantom and brain imaging data show that the proposed method outperforms the state-of-the-art methods in terms of suppressing artifacts and achieving the lowest error. Moreover, the proposed method exhibits robust reconstruction even when the auto-calibration signals are limited in parallel imaging. Overall the proposed method can be exploited to achieve better image quality for accelerated parallel magnetic resonance imaging.
\end{abstract}


\IEEEpeerreviewmaketitle
\section{Introduction}\label{Section:introduction}
Magnetic resonance imaging (MRI) serves as an indispensable tool in clinical diagnosis, but suffers from relatively long data acquisition time \cite{2017_review_PI}. To enable fast imaging, parallel imaging and sparse sampling are two featured approaches. The former is armed with multi coils in the equipment while the latter breaks the Nyquist sampling barrier for sparse images.

\begin{figure}[htbp]
\setlength{\abovecaptionskip}{0pt}
\setlength{\belowcaptionskip}{0pt}
\centering
\includegraphics[width=3.2in]{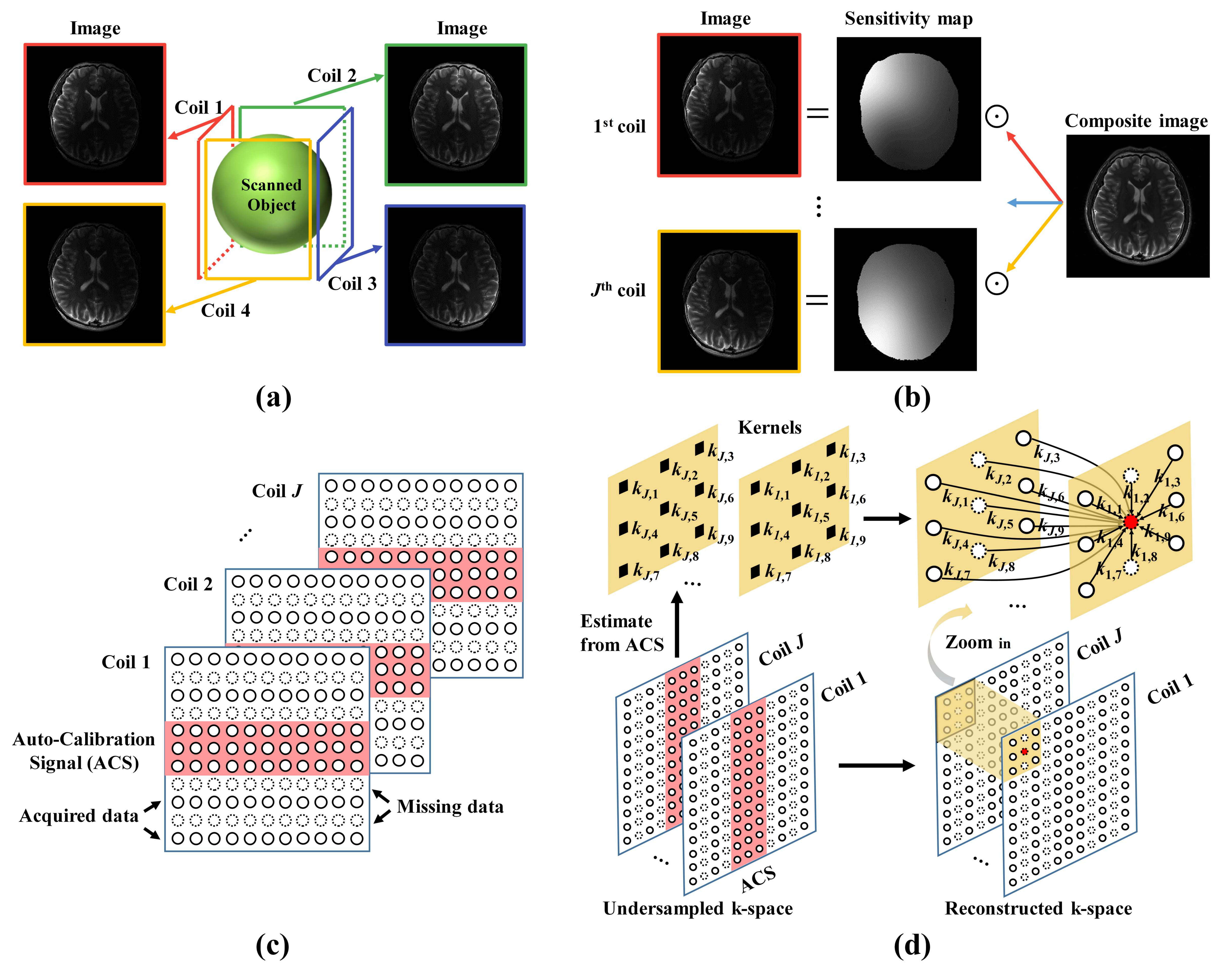}
\caption{Parallel imaging and reconstruction methods. (a) Diagram of parallel imaging; (b) SENSE; (c) auto-calibration signal; (d) SPIRiT.}
\label{fig_PI_SENSE_SPIRiT}
\end{figure}

In parallel imaging, reduced k-space data are acquired using an array of receiver coils (Fig. \ref{fig_PI_SENSE_SPIRiT} (a)) to accelerate data acquisition. Each receiver coil is more sensitive to the specific area of the object near the coil, which allows acquired data to contain more information than single-coil data do. However, undersampling of k-space data results in aliased images, thus reconstruction methods are required to recover missing data for the sake of providing artifact-free images. These parallel imaging reconstruction approaches could be categorized into two main genres: Image domain methods which base on the assumption that there exists a composite image and images of coils can be obtained by multiplying composite image with corresponding sensitivity maps (Fig. \ref{fig_PI_SENSE_SPIRiT} (b)), such as sensitivity encoding (SENSE) \cite{1999_SENSE}. Another type is the k-space domain method of which lies deeply on the fact that each k-space data points of a given coil can be formulated as a linear combination of the multi-coil signals of its neighboring k-space points (Fig. \ref{fig_PI_SENSE_SPIRiT} (d)), such as generalized autocalibrating partially parallel acquisitions (GRAPPA) \cite{2002_GRAPPA}, iterative self-consistent parallel imaging reconstruction (SPIRiT) \cite{2010_SPIRiT} and so on. These parallel imaging approaches achieve reliable reconstruction and some of them have been applied to clinical MRI scans. However, they need auto-calibration signals (ACS), as shown in Fig. \ref{fig_PI_SENSE_SPIRiT} (c), to estimate the coil sensitivity maps for encoding, such as SENSE, or to estimate the kernels for recovery of the missing k-space data, such as GRAPPA and SPIRiT. Once the number of acquired ACS is limited, the accuracy of estimated sensitivity maps or kernels decreases, resulting in large reconstruction errors \cite{2014_MRM_Lustig}. Therefore, how to properly reconstruct the image under limited ACS is still challenging.%
\begin{figure*}[htbp]
\setlength{\abovecaptionskip}{0.cm}
\setlength{\belowcaptionskip}{-0.cm}
\centering
\includegraphics[width=6.4in]{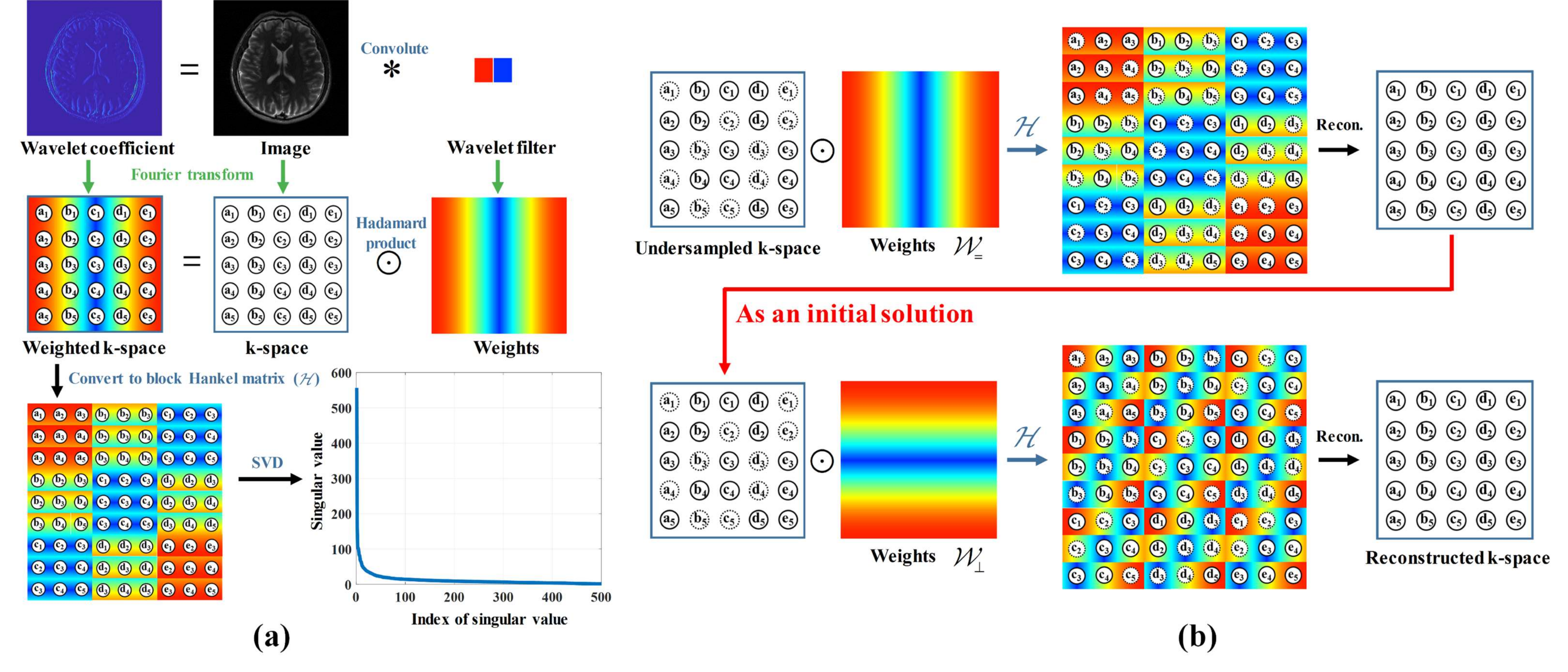}
\caption{Illustration of ALOHA. (a) Low-rankness of weighted k-space with wavelet filters; (b) the sequential image reconstruction for two directions. Note: ${{\mathbf{W}}_{=}}$ and ${{\mathbf{W}}_{\perp}}$ denote the weights in the horizontal and vertical direction.}
\label{fig_ALOHA}
\end{figure*}%

In sparse sampling, k-space data are undersampled and then reconstructed with prior constraints. Two common constraints are sparsity and low-rankness. Sparsity approaches seek a sparse representation of an image under pre-constructed \cite{2007_Sparse_MRI, 2007_MRI_TV, 2017_PCS_EMBC, 2016_pFISTA, 2010_xiaobo,2018_MIA_reweighted_sparse_algorithm} or adaptive \cite{2016_TBME_Zhifang, 2016_MIA_Lai, 2011_Ravishankar, 2015_blind_CS_Ravishankar, 2016_Data_driven_Ravishankar, 2014_PANO, 2013_MRI_xiaobo, 2014_TBME_Wang} basis or dictionaries. Low-rankness methods commonly explore the linear correlations among multiple MRI image \cite{2007_PSF_Liang} \cite{2012_TMI_Zhao} \cite{2015_LR_sparse_MRI}, thus, they are primarily suitable for dynamic or higher-dimensional imaging \cite{2014_PLoSOne_Yu, 2016_high_dim_MR_tensor, 2015_MIA_Zhang,2018_MIA_LRTV_algorithm}. With in-depth and comprehensive developments, the low-rankness of k-space data of a single image or multiple images can be achieved \cite{1989_polynomial_Liang, 2014_MRM_Lustig, 2014_LORAKS, 2016_P_LORAKS, 2016_Off_the_grid, 2016_ALOHA}, which has produced promising image reconstructions under more flexible undersampling patterns, e.g. both randomly and uniformly sparse sampling. Rather than treating the whole k-space as a column of a matrix, these new approaches \cite{1989_polynomial_Liang, 2014_MRM_Lustig, 2014_LORAKS, 2016_P_LORAKS, 2016_Off_the_grid, 2016_ALOHA} have transformed the k-space into structured matrices, e.g. Hankel or Toeplitz matrices, to mine the linear correlation.%

Researchers explore the low-rankness of structured matrices from different properties, e.g. coil sensitivities have compact k-space support in SAKE \cite{2014_MRM_Lustig}, the image has limited spatial support or slowly varying phase in LORAKS \cite{2014_LORAKS,2015_AC_LORAKS,2015_W_LORAKS,2016_P_LORAKS}, transform (wavelets or derivatives) domain sparsity in the ALOHA \cite{2016_ALOHA}, Ongie’s method \cite{2016_Off_the_grid} and GSLR \cite{2018_Hu_LR_TMI}. These low rank structured matrix approaches obviously improve the image reconstruction and provide flexibility to both uniformly and randomly sparse sampling. Among them, those taking advantage of the low-rank property, derived from the transform domain sparsity, of weighted Hankel matrix in k-space hold the flexibility to retain low-rankness since MRI images can be sparsely represented via various transforms. ALOHA \cite{2016_ALOHA} grew out of this observation and proposed to utilize the low-rankness of weighted k-space data (Fig. 2) in reconstruction. However, ALOHA separately enforces the low-rankness of vertical and horizontal directions (Fig. 2 (b)), which results in ALOHA yielding the suboptimal solution.
Ongie et al. \cite{2016_Off_the_grid} and GSLR \cite{2018_Hu_LR_TMI} proposed to simultaneously enforce the low-rankness of both vertical and horizont,al directions. However, though calibration data, i.e. ACS (Fig. 1 (c)), is common in parallel imaging and these low-rank structured matrix methods \cite{2015_AC_LORAKS, 2016_ALOHA,2016_P_LORAKS,2018_Hu_LR_TMI} are compatible to the parallel imaging as shown in TABLE \ref{Table 1}, none of them take the calibration data into consideration for better image reconstruction. It should be noticed that AC-LORAKS \cite{2015_AC_LORAKS} only used the calibration data for algorithm acceleration but not for performance enhancement.%

\begin{table}[htbp]
  \footnotesize
  \centering
  \caption{Differences among Low Rank Structured Matrix Methods}
  \label{Table 1}
    \begin{tabular}{m{1.9cm}<{\centering}m{1cm}<{\centering}m{1.2cm}<{\centering}m{1.3cm}<{\centering}m{1.4cm}<{\centering}}
    \hline
    Method & Parallel Imaging  & k-space Weighting   & Simultaneous Weighting & Calibration Consistency  \\
    \hline
    SAKE         & \checkmark     & $\times$       & /            & $\times$  \\
    LORAKS family& \checkmark     & \checkmark     & \checkmark   & \checkmark  \\
    GSLR         & \checkmark     & \checkmark     & \checkmark   & $\times$ \\
    ALOHA        & \checkmark     & \checkmark     & $\times$     & $\times$  \\
    \hline
    STDLR-SPIRiT & \checkmark     & \checkmark     & \checkmark   & \checkmark \\
    \hline
    \end{tabular}%
    \begin{tablenotes}[flushleft]
    \footnotesize
    \item Note: k-space weighting means k-space data Hadamard multiplying by weights which are the Fourier transform of the convolved filters in sparse representations. Simultaneous weighting indicates imposing low-rank constraints on horizontal and vertical directions. Calibration consistency means exploiting the data consistency from calibration data. The slash / means the corresponding operation is not needed or implemented yet.
     \end{tablenotes}
\end{table}%

In this work, we develop a k-space domain reconstruction method that jointly explores the low-rankness and calibration data. The two directional sparsity of the target image is simultaneously enforced by minimizing the low-rankness of the two directional weighting of the k-space. Besides, intra- and inter-coil data relationships are enforced in the form of self-consistency over k-space. Overall, it is expected that the new approach can further reduce the image reconstruction errors in parallel MRI.

The remainder of the paper is organized as follows. In Section \ref{Section:relatedWork}, we briefly review the related $\ell_1$-SPIRiT and ALOHA methods. In Section \ref{Section:proposedMethod}, the proposed method and algorithm are described. Section \ref{Section:experimentalResults} demonstrates the reconstruction performance and Section \ref{Section:discussions} discusses the effects of the number of ACS lines, parameter settings, empirical convergence of algorithm and comparison with other state-of-the-art low-rank structured matrix methods. Finally, conclusions will be drawn in Section \ref{Section:conclusion}.%

\section{Related Work} \label{Section:relatedWork}
\subsection{$\ell_1$-SPIRiT}
SPIRiT reconstructs parallel MRI image by enforcing the self-consistency of multi-coil k-space data \cite{2010_SPIRiT}. As shown in Fig. \ref{fig_PI_SENSE_SPIRiT} (d), SPIRiT first estimates the linear relation of the intra- and inter-coil k-space data from a small area of the fully sampled k-space center, which is also called as ACS (Fig. \ref{fig_PI_SENSE_SPIRiT} (c)), and then applies this relationship to the rest of the k-space data. In other word, SPIRiT enforces calibration consistency between every point in k-space and its entire neighborhood across all coils in an operator form:
\begin{equation}\label{(1)}
     \mathbf{X}=\mathcal{G}\mathbf{X},
\end{equation}%
where $\mathbf{X}$ denotes k-space data for all coils, and $\mathcal{G}$ is an operator that convolves the k-space data with a series of calibration kernels that are estimated from the ACS (Fig. \ref{fig_PI_SENSE_SPIRiT} (c))\cite{2010_SPIRiT}. Let $\mathbf{Y}$ be the acquired k-space data with non-acquired positions zero-filled for all coils and $\mathcal{U}$ represents the operator that performs undersampling and zerofilling on non-acquired data points, the data acquisition was given by:
\begin{equation}\label{(2)}
    \mathbf{Y}=\mathcal{U}\mathbf{X}.
\end{equation}%

Then, the SPIRiT reconstruction was formulated as:
\begin{equation}\label{(3)}
	\underset{\mathbf{X}}{\mathop{\min }} \, \left\| \mathcal{G}\mathbf{X}-\mathbf{X} \right\|_{F}^{2} \ \ s.t. \ \ \left\| \mathbf{Y}-\mathcal{U}\mathbf{X} \right\|_{F}^{2}\le \varepsilon,
\end{equation}%
where ${{\left\| \cdot  \right\|}_{F}}$ denotes the Frobenius norm, and $\varepsilon $ is a parameter that constraints the noise in the measurements. It has been observed that SPIRiT outperformed SENSE and GRAPPA with superior image quality \cite{2010_SPIRiT}. To better regularize the image, SPIRiT includes an additional penalty in the objective function and constructed the so-called $\ell_1$-SPIRiT model:
\begin{equation}\label{(4)}
    \resizebox{.9\hsize}{!}{$ \underset{\mathbf{X}}{\mathop{\min }}\,\left\| \mathcal{G}\mathbf{X}-\mathbf{X} \right\|_{F}^{2}+\lambda {{\left\| \Psi {{\mathcal{F}}^{-1}}\mathbf{X} \right\|}_{1}} \ \ s.t.\ \ \left\| \mathbf{Y}-\mathcal{U}\mathbf{X} \right\|_{F}^{2}\le \varepsilon $,}
\end{equation}%
where $\Psi $ denotes the sparse transform, ${{\mathcal{F}}^{-1}}$ denotes the inverse Fourier transform and parameter $\lambda$ balances the $\ell_1$ constraint and calibration consistency.

\subsection{ALOHA}\label{Subsection:proposedMethod}
In ALOHA \cite{2016_ALOHA}, the MRI image ${\mathbf{S}} \in {\mathbb{C}^{M \times N}}$ is assumed to be sparse in transform domain. For instance,
\begin{equation}\label{(5)}
    \Psi {\bf{S}} = {\bf{Z}},
\end{equation}%
where $\Psi $ denotes a sparse transform, such as wavelet, and the sparse signal $\mathbf{Z}$ is modeled as the finite superposition of Dirac functions as
\begin{equation}\label{(6)}
	\mathbf{Z}=\sum\limits_{i=1}^{R}{{{a}_{i}}\delta \left( x-{{x}_{i}},y-{{y}_{i}} \right)},
\end{equation}%
where ${{a}_{i}}$ denotes the amplitude of the $i^{\text{th}}$ Dirac function, ${{x}_{i}}$ and ${{y}_{i}}$ the shifts along the vertical and horizontal axes, and $R$ the number of Dirac functions.

Performing Fourier transform on \eqref{(5)}, and with the convolution theorem of Fourier transform, we obtain the spectral form of $\mathbf{Z}$ as below
\begin{equation}\label{(7)}
    \mathbf{M}=\mathcal{F}\left( \mathbf{Z} \right)=\mathcal{F}\left( \Psi \mathbf{S} \right)=\mathcal{F}\left( \Psi  \right)\odot \mathcal{F}\left( \mathbf{S} \right)=\mathbf{W}\odot \mathcal{F}\left( \mathbf{S} \right),
\end{equation}%
where $\odot $ denotes the Hadamard product and $\mathcal{F}$ denotes the Fourier transform and $\mathbf{W}=\mathcal{F}\left( \Psi  \right)$ denotes the weights obtained from applying Fourier transform to sparse transform filters. (See Fig. \ref{fig_ALOHA} (a) for a schematic illustration).

According to the Theorem 2.1 in \cite{2016_ALOHA}, the weighted k-space $\mathbf{M}$ satisfies the low-rankness of the Hankel matrix $\mathcal{H}\mathbf{M}$ as
\begin{equation}\label{(8)}
	rank\left( \mathcal{H}\mathbf{M} \right)=R.
\end{equation}%
where $\mathcal{H}$ is an operator that converts $\mathbf{M}$ into a block Hankel matrix \cite{1992_matrixPencil_Hua, 2014_LR_Yuejie,2018_TBME_Hengfa,2017_ACCESS_Guo} with a dimension of ${({{k}_{1}}{{k}_{2}})\times ((N-{{k}_{1}}+1)(M-{{k}_{2}}+1))}$, and ${{k}_{1}}$ and ${{k}_{2}}$ are pencil parameters.

Assuming that the image is sparse in the transform domain, then very few nonzeros are presented in $\mathbf{Z}$, leading to an effective $R$ be much smaller than the size of block Hankel matrix $\mathcal{H}\mathbf{M}$. Therefore, $\mathcal{H}\mathbf{M}$ enjoys low-rank properties. Ultimately, the recovery of undersampled k-space data can be regarded as a low rank matrix completion problem \cite{2016_ALOHA} as:
\begin{equation}\label{(9)}
	\underset{\mathbf{M}}{\mathop{\min }}\,{{\left\| \mathcal{H}\mathbf{M} \right\|}_{*}} \ \ s.t.\ \ \mathcal{U}\mathbf{M}=\mathbf{\hat{Y}},
\end{equation}%
where $\mathbf{\hat{Y}}=\mathbf{W}\odot \mathbf{Y}_{s}$, $\mathbf{Y}_{s}$ denotes the acquired k-space of a single coil and ${{\left\| \cdot  \right\|}_{*}}$ denotes the matrix nuclear norm.

Extending the above model to multi-coil data acquisition in parallel MRI, the image ${{\mathbf{S}}_{j}}$ of the $j^{\text{th}}$ coil is represented as:
\begin{equation}\label{(10)}
	{{\mathbf{S}}_{j}}={{\mathbf{C}}_{j}}\odot \mathbf{S}_{comp},\ j=1,2,...,J,
\end{equation}%
where ${{\mathbf{C}}_{j}}$ denotes the sensitivity map of the $j^{\text{th}}$ coil, $\mathbf{S}_{comp}$ the composite image.

A fatter matrix cascaded of Hankel matrices is formed as below
\begin{equation}\label{(11)}
    \resizebox{.8\hsize}{!}{$\mathbf{L}=\left[ \mathcal{H}\left( \mathbf{W}\odot {{\mathbf{X}}_{1}} \right),...,\mathcal{H}\left( \mathbf{W}\odot {{\mathbf{X}}_{j}} \right),...\mathcal{H}\left( \mathbf{W}\odot {{\mathbf{X}}_{J}} \right) \right]$},	
\end{equation}%
where ${{\mathbf{X}}_{j}}=\mathcal{F}\left( \mathbf{S}_{j} \right)$ denotes the k-space data of the $j^{\text{th}}$ coil. According to the derivation in \cite{2016_ALOHA}, the matrix $\mathbf{L}$ is low-ranked. Then, the ALOHA for parallel MRI reconstruction \cite{2016_ALOHA} was expressed as:
\begin{equation}\label{(12)}
	\begin{matrix}
  \underset{{{\mathbf{M}}_{j}}}{\mathop{\min }}\,{{\left\| \left[ \mathcal{H}{{\mathbf{M}}_{1}},....\mathcal{H}{{\mathbf{M}}_{J}} \right] \right\|}_{*}}\ , \\
  s.t.\ \mathcal{U}{{\mathbf{M}}_{j}}=\mathbf{W}\odot {{\mathbf{Y}}_{j}},\ j=1,...,J. \\
\end{matrix}
\end{equation}%
where ${{\mathbf{M}}_{j}}=\mathbf{W}\odot {{\mathbf{X}}_{j}}$ and ${{\mathbf{Y}}_{j}}$ denotes the acquired k-space data in $j^{\text{th}}$ coil with non-acquired positions zero-filled.

It's important to note that when applying weighting on 2D k-space data, ALOHA \cite{2016_ALOHA} applies weighting sequentially on each direction. Specifically speaking, it first deals with the optimization problem \eqref{(13)} that enforces weighting along horizontal direction (Fig. \ref{fig_ALOHA} (b)), and then the solution of \eqref{(13)} will be used as an initialization input for the optimization problem \eqref{(14)} which applies weighting along vertical direction (Fig. \ref{fig_ALOHA} (b)).
\begin{equation}\label{(13)}
  \underset{{{\mathbf{M}}^{=}}}{\mathop{\min }}\,{{\left\| \mathcal{H}{{\mathbf{M}}^{=}} \right\|}_{*}} \ \ s.t.\ \ \mathcal{U}{{\mathbf{M}}^{=}}={{\mathbf{W}}_{=}}\odot \mathbf{Y}.
\end{equation}%
\begin{equation}\label{(14)}
  \underset{{{\mathbf{M}}^{\perp}}}{\mathop{\min }}\,{{\left\| \mathcal{H}{{\mathbf{M}}^{\perp}} \right\|}_{*}} \ \ s.t.\ \ \mathcal{U}{{\mathbf{M}}^{\perp}}={{\mathbf{W}}_{\perp}}\odot \mathbf{Y}.
\end{equation}%
where ${{\mathbf{W}}_{=}}$ and ${{\mathbf{W}}_{\perp}}$ denote the weights which are the Fourier transform of filters in the horizontal and vertical directions (Fig. \ref{fig_ALOHA}). Please note that ALOHA replaces the nuclear norm terms in \eqref{(13)} and \eqref{(14)} with the equivalent non-convex terms using the matrix factorizations technique \cite{2004_matrix_factorizations,2017_ACCESS_Guo,2018_TBME_Hengfa} for algorithm acceleration when ALOHA implementation.

\begin{figure}[!h]
\setlength{\abovecaptionskip}{0.cm}
\setlength{\belowcaptionskip}{-0.cm}
\centering
\includegraphics[width=2.8in]{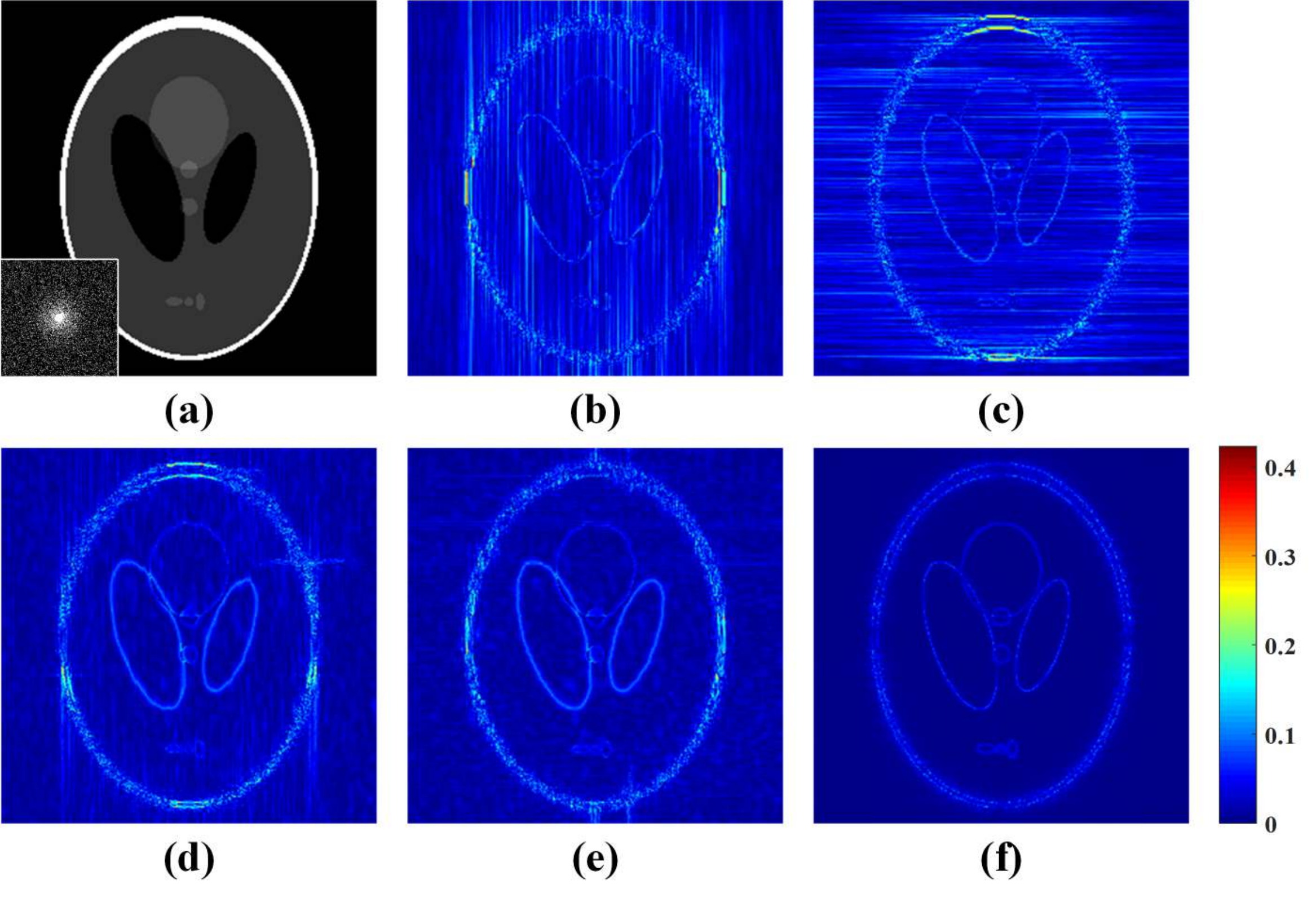}
\caption{Reconstructed single-coil Shepp-Logan phantom using STDLR and ALOHA. (a) Fully sampled Shepp-Logan phantom and 2D random sampling pattern at sampling rate 0.15; (b)-(e) are reconstruction errors using ALOHA and (f) is the reconstruction error using STDLR. (b) (or (c)) is the reconstruction error produced by solely solving sub-problem (\ref{(13)}) (or (\ref{(14)})), respectively; (d) (or (e)) is the reconstruction error by first solving sub-problem (\ref{(13)}) and then (\ref{(14)}) (or first solving (\ref{(14)}) and then (\ref{(13)})) sequentially. Note: The ALOHA with one-level pyramidal decomposition (determined by the filter size) was compared here.}
\label{fig_STDLR_ALOHA}
\end{figure}

The non-convexity of the optimization problems and the sequential processing of the two non-convex problems, however, make ALOHA undergoes the scarce of low rank constraint along both directions simultaneously, resulting in the sub-optimal reconstruction (Fig. \ref{fig_STDLR_ALOHA}).
As shown in Figs. \ref{fig_STDLR_ALOHA} (b) and (c), distinct reconstruction errors are presented when solving (\ref{(13)}) (or (\ref{(14)})) as the first step for sequential reconstruction. These errors may propagate into the subsequent reconstruction and lead to a significant difference in the final reconstructed images (Figs. \ref{fig_STDLR_ALOHA} (d) and (e)). In addition, ALOHA provides obviously different solutions when changing the order of solving two sub-problems (Figs. \ref{fig_STDLR_ALOHA} (d) and (e)), leading to difficulty in choosing the optimal order in practical applications.

Therefore, properly exploring the low-rankness in k-space along horizontal and vertical directions is very important to reduce the image reconstruction errors.


\section{Proposed Method} \label{Section:proposedMethod}
\subsection{Proposed reconstruction model}
In this work, we first propose to simultaneously enforce the low-rankness of k-space both along horizontal and vertical directions as follows:
\begin{equation}\label{(15)}
    \small{\left (\mathbf{STDLR}\right )} \ \ \mathop {\min }\limits_{{{\bf{X}}_s}} {\left\| {{\cal H}{{\bf{W}}_ = }{{\bf{X}}_s}} \right\|_*} + {\left\| {{\cal H}{{\bf{W}}_ \perp }{{\bf{X}}_s}} \right\|_*} + \frac{\lambda }{2}\left\| {{{\bf{Y}}_s} - \mathcal{U}{{\bf{X}}_s}} \right\|_F^2,
\end{equation}%
where $\mathbf{X}_s$ denotes the targeted single-coil k-space data, $\mathbf{Y}_s$ the acquired single-coil k-space data and the nuclear norm term ${{\left\| \cdot  \right\|}_{*}}$ imposes the low-rank constraint on the weighted Hankel structured matrix. We call this model, one of the two proposed models in this work, STDLR for short. Please note that the way we enforce jointly weighting is different from the way Ongie's \cite{2016_Off_the_grid} and GSLR methods \cite{2018_Hu_LR_TMI} did in which they cascaded the two weighting matrices together to enforce low-rank constraint. In our way to enforce simultaneous weighting, the low-rank matrices lie in much smaller signal space expected to alleviate computation complexity. Compared with ALOHA, STDLR imposes a stronger low rank constraint allowing a possible better reconstruction. Reconstructions on Shepp-Logan phantom (Fig. \ref{fig_STDLR_ALOHA}) show that STDLR achieves lower reconstruction error than ALOHA. 

A basic way to extend STDLR into parallel MRI is cascading the Hankel matrix of each coil into a fatter matrix \cite{2016_P_LORAKS,2016_ALOHA} as:
\begin{equation}\label{HWXH}
\begin{aligned}
    \tilde{\mathcal{H}}{{\mathcal{W}}_{=}}\mathbf{X} & =\left[ \mathcal{H}{{\mathbf{W}}_{=}}\odot {{\mathbf{X}}_{1}},...,\mathcal{H}{{\mathbf{W}}_{=}}\odot {{\mathbf{X}}_{J}} \right],\\
     \tilde{\mathcal{H}}{{\mathcal{W}}_{\perp}}\mathbf{X} & =\left[ \mathcal{H}{{\mathbf{W}}_{\perp}}\odot {{\mathbf{X}}_{1}},...,\mathcal{H}{{\mathbf{W}}_{\perp}}\odot {{\mathbf{X}}_{J}} \right],\\
\end{aligned}
\end{equation}%
where  $\mathbf{X}=\left[ {{\mathbf{X}}_{1}},{{\mathbf{X}}_{2}},...,{{\mathbf{X}}_{J}} \right]$ denotes the targeted k-space data from all coils and $\mathbf{Y}=\left[ {{\mathbf{Y}}_{1}},{{\mathbf{Y}}_{2}},...,{{\mathbf{Y}}_{J}} \right]$ are acquired k-space data with zero-filling at non-acquired positions, and ${{\mathcal{W}}_{=}}\mathbf{X}=\left[ {{\mathbf{W}}_{=}}\odot{{\mathbf{X}}_{1}},...,{{\mathbf{W}}_{\perp}}\odot {{\mathbf{X}}_{J}} \right]$,
${{\mathcal{W}}_{\perp}}\mathbf{X}=\left[ {{\mathbf{W}}_{\perp}}\odot{{\mathbf{X}}_{1}},...,{{\mathbf{W}}_{\perp}}\odot {{\mathbf{X}}_{J}} \right]$,

Then, a basic STDLR with cascaded Hankel matrices can be modeled as:
\begin{equation}\label{(15-1)}
    \mathop {\min }\limits_{\bf{X}} {\left\| {\tilde {\cal H}{{\cal W}_= }{\bf{X}}} \right\|_*} + {\left\| {\tilde {\cal H}{{\cal W}_ \perp}{\bf{X}}} \right\|_*} + \frac{\lambda }{2}\left\| {{\bf{Y}} - {\cal U}{\bf{X}}} \right\|_F^2.
\end{equation}%
Although the cascaded Hankel matrices were designed to implicitly utilize the correlation among multiple coils, we found this approach still suboptimal (Fig. \ref{fig_STDLR_SPIRiT_advan}(a)).

\begin{figure}[htbp]
\setlength{\abovecaptionskip}{0.cm}
\setlength{\belowcaptionskip}{-0.cm}
\centering
\includegraphics[width=3.4in]{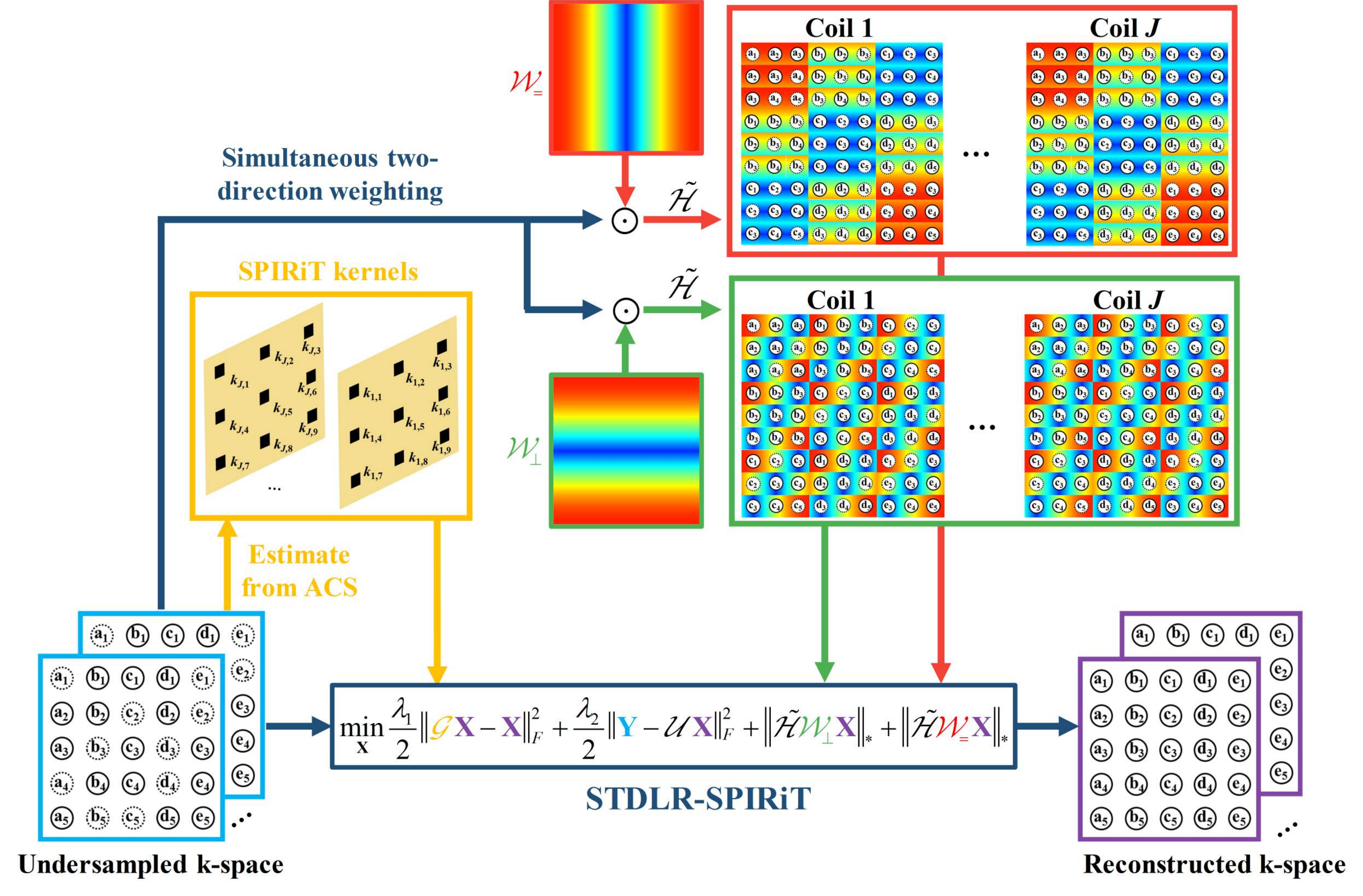}
\caption{The flowchart of the proposed method. Note: ${{\mathbf{W}}_{=}}$ and ${{\mathbf{W}}_{\perp}}$ denote the weights in the horizontal and vertical direction.}
\label{fig_flowchart_STDLR_SPIRiT}
\end{figure}

\begin{figure}[htbp]
\setlength{\abovecaptionskip}{0.cm}
\setlength{\belowcaptionskip}{-0.cm}
\centering
\includegraphics[width=2.8in]{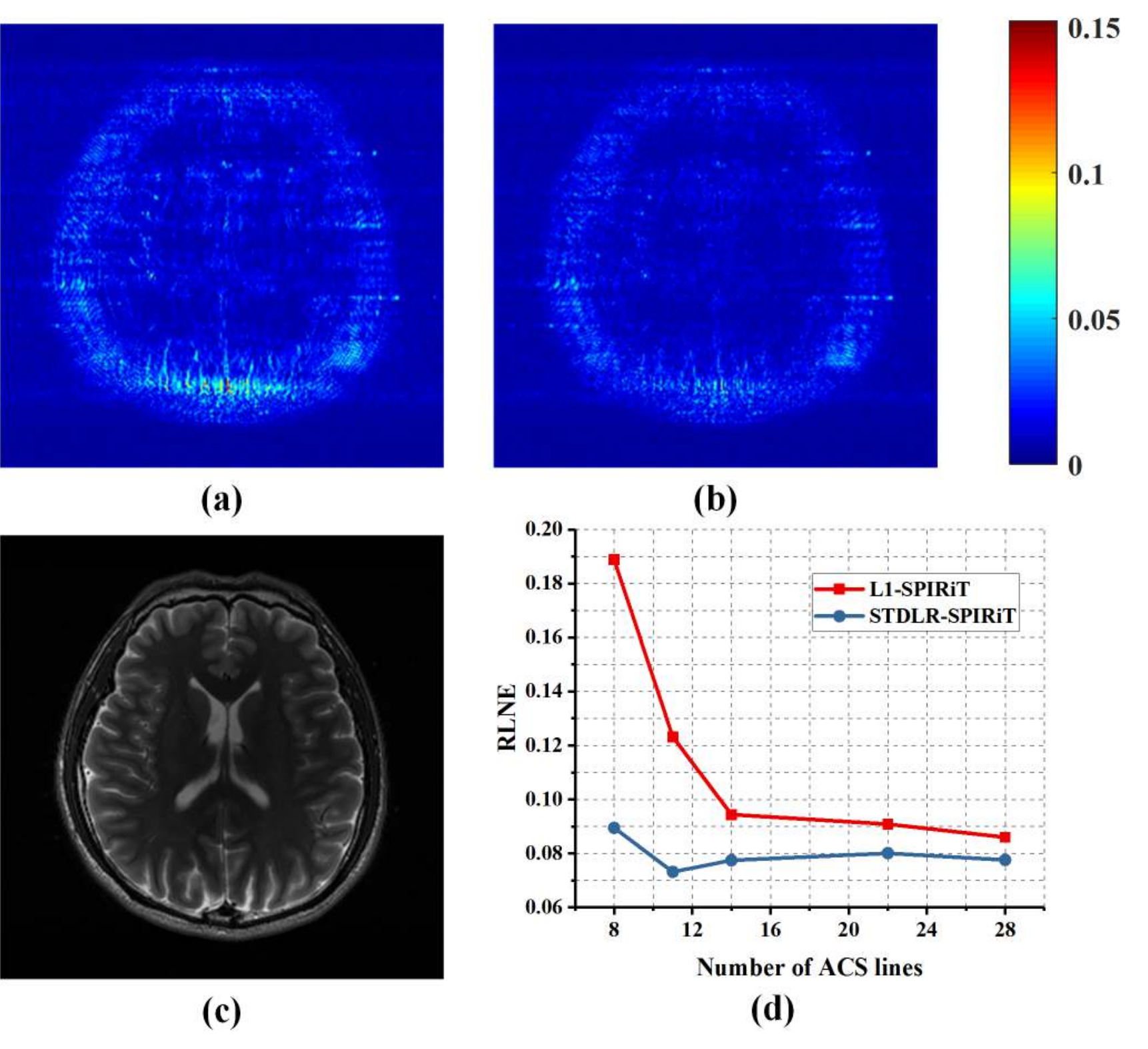}
\caption{Reconstruction merits of 4-coil brain image using STDLR-SPIRiT. (a)-(b) are the reconstruction errors of STDLR with cascaded Hankel matrices and STDLR-SPIRiT; (c) a square root of sum of squares image of fully sampled data; (d) is the
reconstruction errors versus number of ACS lines by $\ell_1$-SPIRiT and STDLR-SPIRiT. All reconstructions in this experiment are conducted under 1D Gaussian Cartesian undersampling pattern with a fixed sampling rate of 0.34.}
\label{fig_STDLR_SPIRiT_advan}
\end{figure}

To reinforce the reconstruction in parallel imaging, our intention is to mine the self-consistency, that are estimated from the calibration data, over multi-coil k-space. In parallel MRI, SPIRiT achieved promising results and suggested a valuable concept that each k-space signal of a specific coil can be formulated as a linear combination of the multi-coil signals of its neighboring k-space points \cite{2002_GRAPPA, 2010_SPIRiT}. Taking this self-consistency property into account, we further exploit the aforementioned proposed STDLR model in \eqref{(15)} to cope with parallel imaging as:
\begin{equation} \label{(16)}
    \begin{aligned}
        \small{\left ( \textbf{STDLR-SPIRiT} \right )} \quad & \mathop {\min }\limits_{\bf{X}} {\left\| {\tilde {\cal H}{{\cal W}_ \bot }{\bf{X}}} \right\|_*} + {\left\| {\tilde {\cal H}{{\cal W}_ = }{\bf{X}}} \right\|_*} + \\
        & \frac{{{\lambda _1}}}{2} \left\| {{\cal G}{\bf{X}} - {\bf{X}}} \right\|_F^2 + \frac{{{\lambda _2}}}{2}\left\| {{\bf{Y}} - {\cal U}{\bf{X}}} \right\|_F^2, \\
    \end{aligned}
\end{equation}%
where ${{\lambda}_{1}}$ and ${{\lambda }_{2}}$ trade off among the low-rank constraint, self-calibration consistency and undersampled data fidelity. The second proposed approach uses the same calibration $\mathcal{G}$ as the SPIRiT. Since the third term is the same as the SPIRiT, we call the second proposed model in \eqref{(16)} as STDLR-SPIRiT. The schematic illustration of STDLR-SPIRiT is depicted in Fig. \ref{fig_flowchart_STDLR_SPIRiT}. The STDLR-SPIRiT holds the advantage over the STDLR with cascaded Hankel matrices on reducing reconstruction errors at edges. The reason is that STDLR-SPIRiT utilizes the local linear relations existed in intra- and inter-coils that have been explicitly estimated from ACS data (Figs. \ref{fig_STDLR_SPIRiT_advan}(a) and (b)). Besides, although with the same ACS strategy, STDLR-SPIRiT is particularly useful when the ACS data are limited, compared with the state-of-the-art sparsity-based SPIRiT \cite{2010_SPIRiT} as shown in Fig. \ref{fig_STDLR_SPIRiT_advan} (d). These observations imply that both the low-rankness and estimated self-consistency improve the reconstruction in the proposed STDLR-SPIRiT.


To tackle \eqref{(16)}, time-consuming singular value decomposition (SVD) is hardly avoided since singular value thresholding is utilized for nuclear norm minimization \cite{2010_SVT, 2017_ACCESS_Guo, 2018_TBME_Hengfa}. In our case, the size of low-rank matrix cascaded of Hankel matrices $(J{{k}_{1}}{{k}_{2}})\times ((N-{{k}_{1}}+1)(M-{{k}_{2}}+1))$ increases dramatically after pencil parameters ${{k}_{1}},\ {{k}_{2}}$ increase, resulting in a sharp increase in SVD calculation time. For example, the dimension of $\tilde{\mathcal{H}}{{\mathcal{W}}_{=}}\mathbf{X}$ reaches $54756\times 2116$ for a 4-coil $256\times 256$ image with pencil parameters ${{k}_{1}}=23$, ${{k}_{2}}=23$. Performing one-time SVD on $\tilde{\mathcal{H}}{{\mathcal{W}}_{=}}\mathbf{X}$ needs 12.25 second on a CentOS 7 computation server with two Intel Xeon CPUs of 3.5GHz and 112GB RAM. Thus, we seek an SVD-free algorithm to reduce the computation time. Here, we employ matrix factorization technique \cite{2010_SVT, 2017_ACCESS_Guo, 2018_TBME_Hengfa}, leading to a dramatic reduction of computation time. The SVD-free algorithm is based on the relationship \cite{2004_matrix_factorizations} shown below:
\begin{equation}\label{(17)}
    {{\left\| \mathbf{A} \right\|}_{*}}=\underset{\mathbf{P},\mathbf{Q}}{\mathop{\min }}\,\frac{1}{2}\left( \left\| \mathbf{P} \right\|_{F}^{2}+\left\| \mathbf{Q} \right\|_{F}^{2} \right)\  \ s.t.\ \ \mathbf{P}{{\mathbf{Q}}^{H}}=\mathbf{A},
\end{equation}%
where $\mathbf{P}$ and $\mathbf{Q}$ denote two factorized matrices, and the upper subscript $H$ denotes the Hermitian transpose of a complex matrix. Hence, \eqref{(16)} can be reformulated as matrix factorization constraint:
\begin{equation}\label{(18)}
\begin{aligned}
        \underset{\mathbf{X},\mathbf{P},\mathbf{Q}}{\mathop{\min }} & \frac{1}{2}\sum\limits_{i}{\left( \left\| {{\mathbf{P}}_{i}} \right\|_{F}^{2} + \left\| {{\mathbf{Q}}_{i}} \right\|_{F}^{2} \right)}+\frac{{{\lambda }_{1}}}{2}\left\| \mathcal{G}\mathbf{X} - \mathbf{X} \right\|_{F}^{2} + \\
        & \frac{{{\lambda }_{2}}}{2}\left\| \mathbf{Y} - \mathcal{U}\mathbf{X} \right\|_{F}^{2} \ \ s.t. \ \  {{\mathbf{P}}_{i}}\mathbf{Q}_{i}^{H}=\tilde{\mathcal{H}}{{\mathcal{W}}_{i}}\mathbf{X}. \\
\end{aligned}
\end{equation}%
where $i$ denotes $\perp$ or $=$, i.e. vertical and horizontal directions. 

In summary, STDLR-SPIRiT in \eqref{(16)} is the final model of this work. And \eqref{(18)} is the accelerated version of the proposed STDLR-SPIRiT model to avoid the time-consuming SVD computation.

\subsection{SVD-free low rank reconstruction algorithm}
\begin{figure*}[htbp]
\setlength{\abovecaptionskip}{0.cm}
\setlength{\belowcaptionskip}{-0.cm}
\centering
\includegraphics[width=6.4in]{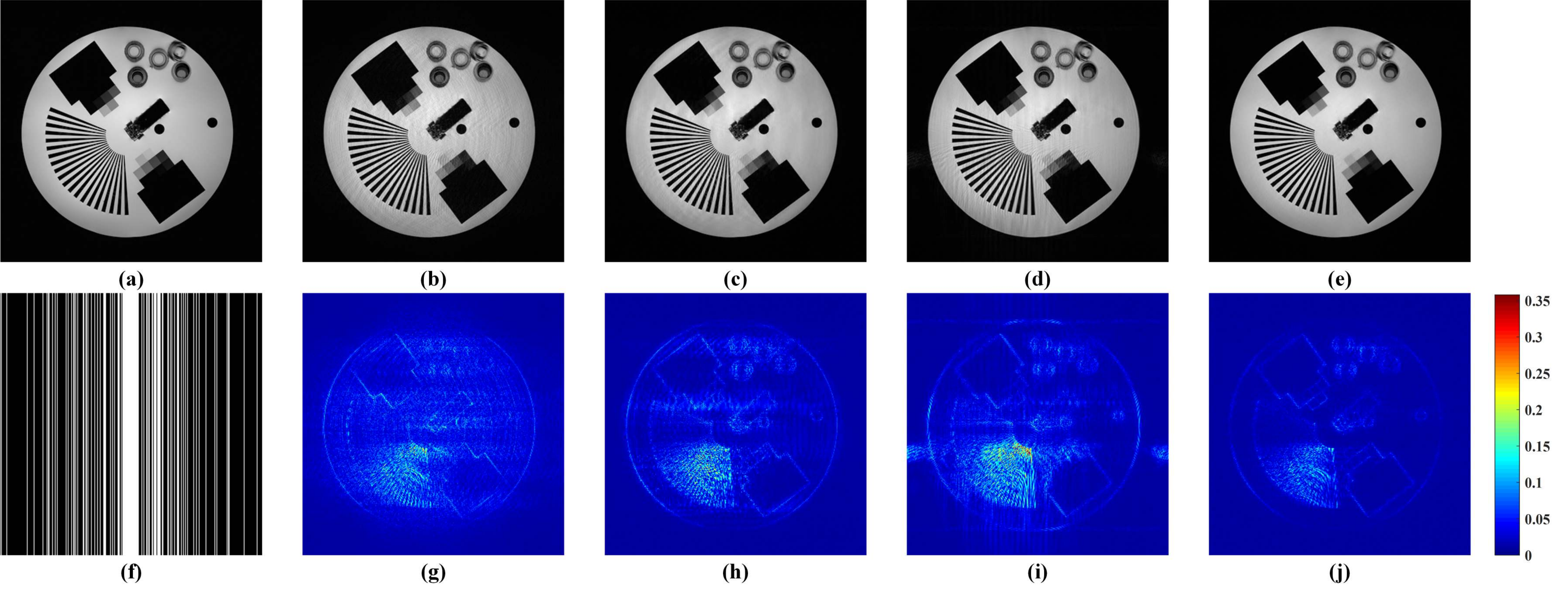}
\caption{Reconstruction results and errors of a phantom using the Cartesian pattern with a sampling rate of 0.27. (a) An SSOS image of fully sampled data; (b-e) SSOS images of reconstructed results by GRAPPA, $\ell_1$-SPIRiT, ALOHA and STDLR-SPIRiT, respectively; (f) the Cartesian undersampling pattern; (g-j) the reconstruction error distribution corresponding to the above methods.}
\label{fig_results_phantom}
\end{figure*}
Here, we adopt the Alternating Direction Method of Multiplier (ADMM) \cite{2011_ADMM} to deal with the proposed model in \eqref{(18)}. It should be noticed that the non-convex optimization problem \eqref{(18)} can be successfully solved by ADMM \cite{2016_ALOHA,2018_TBME_Hengfa,2017_ACCESS_Guo}.
The augmented Lagrangian form of \eqref{(18)} is
\begin{equation}\label{(19)}
    \begin{aligned}
         & \underset{{{\mathbf{D}}_{i}}}{\mathop{\max }} \ \underset{\mathbf{X},{{\mathbf{P}}_{i}},{{\mathbf{Q}}_{i}}}{\mathop{\min }} \frac{{{\lambda }_{1}}}{2}\left\| \mathcal{G}\mathbf{X}-\mathbf{X} \right\|_{F}^{2} + \frac{{{\lambda }_{2}}}{2}\left\| \mathbf{Y}-\mathcal{U}\mathbf{X} \right\|_{F}^{2} +\\
         & \frac{1}{2}\sum\limits_{i} \! {\left( \! \left\| {{\mathbf{P}}_{i}} \right\|_{F}^{2}+\left\| {{\mathbf{Q}}_{i}} \right\|_{F}^{2} \! \right)} \!\! + \!\! \sum\limits_{i} \!{\left\langle \! {{\mathbf{D}}_{i}},\tilde{\mathcal{H}}{{\mathcal{W}}_{i}}\mathbf{X}\!-\!{{\mathbf{P}}_{i}}\mathbf{Q}_{i}^{H} \! \right\rangle} \\
         & + \sum\limits_{i}{\frac{{{\beta }_{i}}}{2}\left\| \tilde{\mathcal{H}}{{\mathcal{W}}_{i}}\mathbf{X}-{{\mathbf{P}}_{i}}\mathbf{Q}_{i}^{H} \right\|_{F}^{2}}, \\
    \end{aligned}
\end{equation}%
where ${{\mathbf{D}}_{i}}$ denotes the Lagrangian multiplier, $\left\langle \cdot ,\cdot  \right\rangle $ the inner product in the Hilbert space of complex matrices, defined by $\left\langle \mathbf{A},\mathbf{B} \right\rangle =\Re \left\langle \mathbf{A}\left( : \right),\mathbf{B}\left( : \right) \right\rangle =\Re \left( trace\left( {{\mathbf{A}}^{*}}\mathbf{B} \right) \right)$, $\Re$ denotes the real part, and ${{\beta }_{i}}$ denotes the penalty parameter for the $i^{\text{th}}$ direction. The solution of \eqref{(19)} can be obtained through alternatively solving sub-problems below:
\begin{equation}\label{sub-pro-1}
\begin{medsize}
    \begin{aligned}
        {{\mathbf{X}}^{\left( k+1 \right)}} & = \arg \underset{\mathbf{X}}{\mathop{\min }} \frac{{{\lambda }_{1}}}{2} \left\| \mathcal{G}\mathbf{X} - \mathbf{X} \right\|_{F}^{2} + \sum\limits_{i} \left\langle \mathbf{D}_{i}^{\left( k \right)},\tilde{\mathcal{H}}{{\mathcal{W}}_{i}}\mathbf{X} - \mathbf{P}_{i}^{\left( k \right)} \right. \left. \mathbf{Q}_{i}^{\left( k \right)H}  \right\rangle \\
        & + \frac{{{\lambda }_{2}}}{2} \left\| \mathbf{Y}-\mathcal{U}\mathbf{X} \right\|_{F}^{2} + \sum\limits_{i} {\frac{{{\beta }_{i}}}{2} \left\|  \tilde{\mathcal{H}}{{\mathcal{W}}_{i}}\mathbf{X} - \mathbf{P}_{i}^{\left( k \right)}\mathbf{Q}_{i}^{\left( k \right)H} \right\|_{F}^{2} } ,\\
    \end{aligned}
\end{medsize}
\end{equation}%
\begin{equation}\label{sub-pro-2}
\begin{medsize}
    \begin{aligned}
        \mathbf{P}_{i}^{\left( k+1 \right)} = & \arg \underset{{{\mathbf{P}}_{i}}}{\mathop{\min }} \left\langle \mathbf{D}_{i}^{\left( k \right)},\tilde{\mathcal{H}}{{\mathcal{W}}_{i}}{{\mathbf{X}}^{\left( k+1 \right)}}-{{\mathbf{P}}_{i}}\mathbf{Q}_{i}^{\left( k \right)H} \right\rangle + \\
        &  \frac{1}{2}\left\| {{\mathbf{P}}_{i}} \right\|_{F}^{2} + \frac{{{\beta }_{i}}}{2}\left\| \tilde{\mathcal{H}}\mathcal{W}{{}_{i}}{{\mathbf{X}}^{\left( k+1 \right)}}-{{\mathbf{P}}_{i}}\mathbf{Q}_{i}^{\left( k \right)H} \right\|_{F}^{2}, \\
    \end{aligned}
\end{medsize}
\end{equation}%
\begin{equation}\label{sub-pro-3}
\begin{medsize}
    \begin{aligned}
        \mathbf{Q}_{i}^{\left( k+1 \right)} = & \arg \underset{{{\mathbf{Q}}_{i}}}{\mathop{\min }} \left\langle \mathbf{D}_{i}^{\left( k \right)},\tilde{\mathcal{H}}{{\mathcal{W}}_{i}}{{\mathbf{X}}^{\left( k+1 \right)}}-\mathbf{P}_{i}^{\left( k+1 \right)}\mathbf{Q}_{i}^{H} \right\rangle + \\
        &  \frac{1}{2}\left\| {{\mathbf{Q}}_{i}} \right\|_{F}^{2} + \frac{{{\beta }_{i}}}{2}\left\| \tilde{\mathcal{H}}{{\mathcal{W}}_{i}}{{\mathbf{X}}^{\left( k+1 \right)}}-\mathbf{P}_{i}^{\left( k+1 \right)}\mathbf{Q}_{i}^{H} \right\|_{F}^{2}, \\
    \end{aligned}
\end{medsize}
\end{equation}%
\begin{equation}\label{sub-pro-4}
\begin{medsize}
    \mathbf{D}_{i}^{\left( k+1 \right)}=\mathbf{D}_{i}^{\left( k \right)}+{{\tau }_{i}}\left( \tilde{\mathcal{H}}{{\mathcal{W}}_{i}}{{\mathbf{X}}^{\left( k+1 \right)}}-\mathbf{P}_{i}^{\left( k+1 \right)}{{\mathbf{Q}}_{i}}^{\left( k+1 \right)H} \right).
\end{medsize}
\end{equation}%
where ${{\tau}_{i}}$ is the step size of the $i^{\text{th}}$ direction and is set as 1 for all $i$. For fixed $\mathbf{P}_{i}^{\left( k \right)}$, $\mathbf{Q}_{i}^{\left( k \right)}$ and $\mathbf{D}_{i}^{\left( k
\right)}$, ${{\mathbf{X}}^{\left( k+1 \right)}}$ has a close-form solution as
\begin{equation}\label{(21)}
\begin{medsize}
    \begin{aligned}
        & {{\mathbf{X}}^{\left( k+1 \right)}}=\bigg( {{\lambda }_{1}}\left( {{\mathcal{G}}^{*}}\mathcal{G}-{{\mathcal{G}}^{*}}-\mathcal{G}+\mathbf{I} \right)+\sum\limits_{i}{{{\beta }_{i}}\mathcal{W}_{i}^{*}{{{\tilde{\mathcal{H}}}}^{*}}\tilde{\mathcal{H}}{{\mathcal{W}}_{i}}} + \\
        &  {{\lambda }_{2}}{{\mathcal{U}}^{*}}\mathcal{U}  \bigg) ^{-1}  \left[ {{\lambda }_{2}}{{\mathcal{U}}^{*}}\mathbf{Y}+\sum\limits_{i}{{{\beta }_{i}}\mathcal{W}_{i}^{*}{{{\tilde{\mathcal{H}}}}^{*}} \left(  \mathbf{P}_{i}^{\left( k \right)}\mathbf{Q}_{i}^{\left( k \right)H}-\frac{\mathbf{D}_{i}^{\left( k \right)}}{{{\beta }_{i}}} \right)} \right] , \\
    \end{aligned}
\end{medsize}
\end{equation}%
where the upper subscript $*$ denotes the adjoint operator.
For fixed ${{\mathbf{X}}^{\left( k+1 \right)}}$, $\mathbf{Q}_{i}^{\left( k \right)}$ and $\mathbf{D}_{i}^{\left( k \right)}$, $\mathbf{P}_{i}^{\left( k+1 \right)}$ is obtained by
\begin{equation}\label{(22)}
\begin{medsize}
\begin{aligned}
	\mathbf{P}_{i}^{\left( k+1 \right)} \!\!=\!\! \left( \! {{\beta }_{i}}\tilde{\mathcal{H}}{{\mathcal{W}}_{i}}{{\mathbf{X}}^{\left( k+1 \right)}}\!+\!\mathbf{D}_{i}^{\left( k \right)} \! \right)
    \! \mathbf{Q}_{i}^{\left( k \right)} \! {{\left( \! \mathbf{I}\!+\!{{\beta }_{i}}\mathbf{Q}_{i}^{\left( k \right)H}\mathbf{Q}_{i}^{\left( k \right)} \! \right)}^{-1}} \!.
\end{aligned}
\end{medsize}
\end{equation}
For fixed ${{\mathbf{X}}^{\left( k+1 \right)}}$, $\mathbf{P}_{i}^{\left( k+1 \right)}$ and $\mathbf{D}_{i}^{\left( k \right)}$, $\mathbf{Q}_{i}^{\left( k+1 \right)}$ can be solved by
\begin{equation}\label{(23)}
\begin{medsize}
\begin{aligned}
	\mathbf{Q}_{i}^{\left( k+1 \right)} = & {{\left( {{\beta }_{i}}\tilde{\mathcal{H}}{{\mathcal{W}}_{i}}{{\mathbf{X}}^{\left( k+1 \right)}} + \mathbf{D}_{i}^{\left( k \right)}  \right)}^{*}} \\
    & \mathbf{P}_{i}^{\left( k+1 \right)} {{\left( \mathbf{I} + {{\beta }_{i}}\mathbf{P}_{i}^{\left( k+1 \right)H}\mathbf{P}_{i}^{\left( k+1 \right)} \right)}^{-1}} . \\
\end{aligned}
\end{medsize}
\end{equation}
The numerical algorithm is summarized in Algorithm \ref{alg:1}.
\begin{algorithm}[htb]
\footnotesize
\caption{MRI image reconstruction with STDLR-SPIRiT}\label{alg:1}
\hspace*{0.02in}{\bf{Input:}}  $\mathbf{Y}$, $\mathcal{U}$, $\mathcal{G}$, ${{\lambda }_{1}}$, ${{\lambda }_{2}}$, ${{\tau }_{i}}$, ${{\beta }_{i}}$ .\\
\hspace*{0.02in}{\bf{Initialization:}} ${{\mathbf{P}}_{i}}$ and ${{\mathbf{Q}}_{i}}$ are initialized as random matrix, ${{\mathbf{D}}_{i}}=\mathbf{1}$, and $k=1$.\\
\hspace*{0.02in}{\bf{Output:}} ${\mathbf{X}}$.
\begin{algorithmic}[1]
\WHILE{ $k\le 100$ and $  \left\| {{\mathbf{X}}^{\left( k+1 \right)}}-{{\mathbf{X}}^{\left( k \right)}}\  \right\|_{F}^{2}\ /\left\| {{\mathbf{X}}^{\left( k \right)}}\  \right\|_{F}^{2}\ge {{10}^{-6}}$ }
\STATE
Update $\mathbf{X}$ by solving equation \eqref{(21)};
\STATE
Update ${{\mathbf{P}}_{i}}$ by using \eqref{(22)};
\STATE
Update ${{\mathbf{Q}}_{i}}$ by using \eqref{(23)};
\STATE
Update multiplier ${{\mathbf{D}}_{i}}$ by using \eqref{sub-pro-4};
\STATE $ k = k + 1 $;
\ENDWHILE
\end{algorithmic}
\end{algorithm}

\begin{table}[htbp]
\footnotesize
  \centering
  \caption{Differences among Compared Methods.}\label{Table 2}
    \begin{tabular}{m{1.95cm}<{\centering}m{1.4cm}<{\centering}m{1cm}<{\centering}m{1.5cm}<{\centering}m{0.9cm}<{\centering}}
    \toprule
    Method & Calibration consistency & Image sparsity & k-space low-rankness & Wavelet \\
    \midrule
    GRAPPA             & \checkmark     & $\times$     & $\times$       & / \\
    $\ell_1$-SPIRiT    & \checkmark     & \checkmark   & $\times$       & db4 \\
    ALOHA              & $\times$       & $\times$     & \checkmark     & Haar \\
    STDLR-SPIRiT       & \checkmark     & $\times$     & \checkmark     & Haar \\
    \bottomrule
    \end{tabular}%
        \begin{tablenotes}[flushleft]
         \footnotesize
         \item Note: If the constraint, such as calibration consistency, image sparsity and k-space low-rankness is checked of the method row, it means the corresponding constraint was enforced directly in the approach.
         Image sparsity means imposing sparse assumption in transform domain. The k-space low-rankness means imposing low rank constraint in k-space domain. Wavelet denotes the type of wavelet used in sparsifying transform or k-space weighting.
     \end{tablenotes}
\end{table}%
\section{Experimental Results}\label{Section:experimentalResults}
In this section, the reconstruction performance of the proposed method is evaluated on phantom and \textit{in vivo} MRI data. Cartesian sampling with random phase encoding, pseudo radial sampling and 2D random sampling data are retrospectively constructed from fully sampled data. The ACS signals of sampling patterns are optimized for kernel estimation. The proposed method, STDLR-SPIRiT, is compared with three state-of-the-art reconstruction methods including GRAPPA, $\ell_1$-SPIRiT and ALOHA. Both GRAPPA and $\ell_1$-SPIRiT are typical parallel imaging reconstruction methods using kernel estimation. In addition, the $\ell_1$-SPIRiT exploits the sparsity of images under wavelets transform. ALOHA with cascaded Hankel matrices in (\ref{(12)}) is chosen for comparison since it utilizes the low-rankness of weighted k-space of multi-coil data. Differences among these methods are summarized in Table \ref{Table 2}.
\begin{figure*}[htbp]
\setlength{\abovecaptionskip}{0.cm}
\setlength{\belowcaptionskip}{-0.cm}
\centering
\includegraphics[width=6.4in]{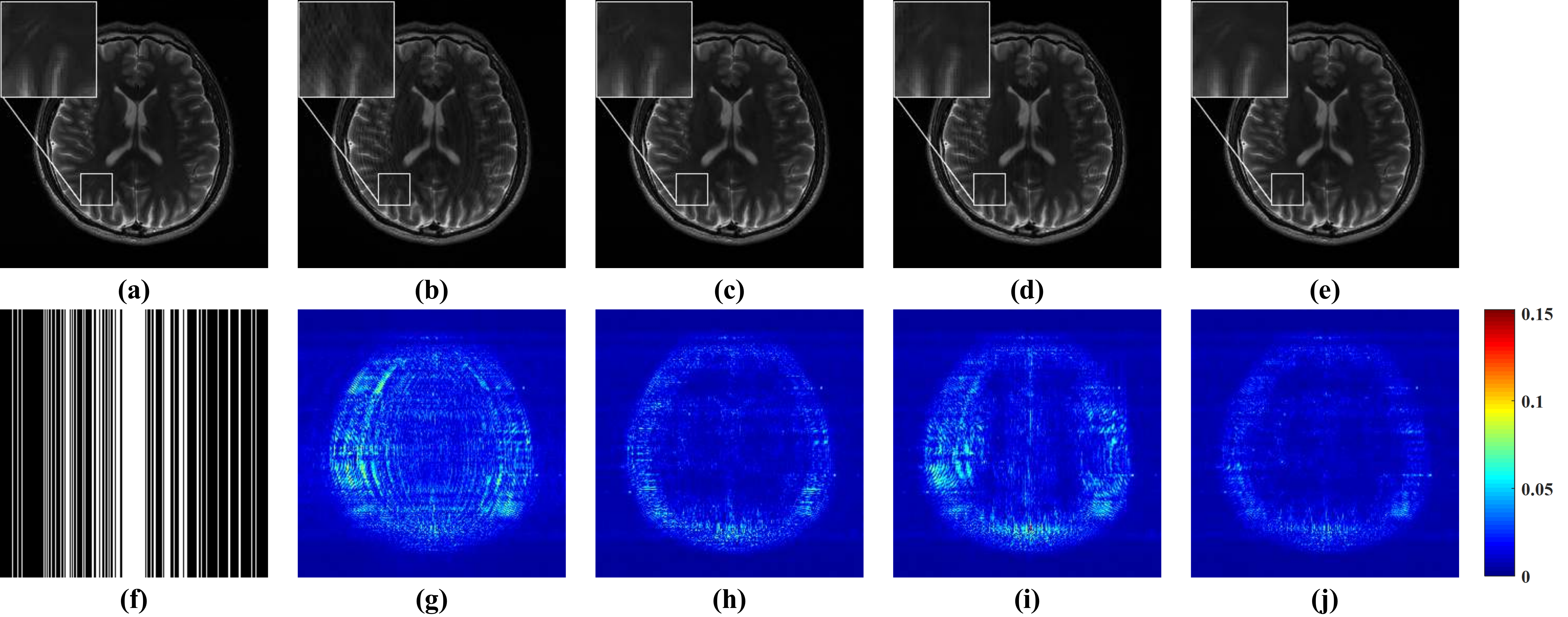}
\caption{Reconstruction results and errors of human brain using the Cartesian pattern with a sampling rate of 0.34. (a) An SSOS image of fully sampled data; (b-e) SSOS images of reconstructed results by GRAPPA, $\ell_1$-SPIRiT, ALOHA and STDLR-SPIRiT, respectively; (f) the Cartesian undersampling pattern; (g-j) the reconstruction error distribution corresponding to reconstructed image above them.}
\label{fig_results_brain_typical}
\end{figure*}

The implementation of GRAPPA and $\ell_1$-SPIRiT was shared at Dr. Michael Lustig’s website \cite{code_SPIRiT} while ALOHA was shared at Dr. Jong Chul Ye’ website \cite{code_ALOHA}. Parameters of all the compared methods are optimized to obtain the lowest RLNE (See Supplementary Material for details). All computation procedures run on a CentOS 7 computation server with two Intel Xeon CPUs of 3.5 GHz and 112 GB RAM. For 4-coil parallel MRI, the reconstruction time of GRAPPA, $\ell_1$-SPIRiT, ALOHA, and the proposed STDLR-SPIRiT are 26.8 seconds, 22.5 seconds, 458.3 seconds, and 1090.3 seconds, respectively.

In all experiments, the reconstructed multi-coil images are combined by a square root of sum of squares (SSOS), and the error distributions of multi coils were combined into single-coil difference image with SSOS.

We adopt relative $\ell_2$ norm error (RLNE) and mean structure similarity index measure (MSSIM) \cite{2004_TIP_Zhou} as objective criteria to quantify the reconstruction performance. The RLNE is defined as
\begin{equation}\label{(24)}
  \text{RLNE}=\frac{{{\left\| \mathbf{x}-\mathbf{\hat{x}} \right\|}_{2}}}{{{\left\| \mathbf{x} \right\|}_{2}}},
\end{equation}%
where $\mathbf{x}$ denotes the column stacked fully sampled k-space data and  $\mathbf{\hat{x}}$ the column stacked reconstructed k-space data. A lower RLNE demonstrates higher consistency between the fully sampled image and the reconstructed image. The mean measure of structural similarity (MSSIM) on two images $\mathbf{A}$ and $\mathbf{B}$ is calculated through
\begin{equation}\label{(25)}
    \resizebox{.8\hsize}{!}{$ \text{MSSIM}\left( \mathbf{A},\mathbf{B} \right) = \frac{1}{M} \sum\limits_{i=1}^{M}{\frac{\left( 2{{\mu }_{{{\mathbf{a}}_{i}}}}{{\mu }_{{{\mathbf{b}}_{i}}}} + {{C}_{1}} \right) \left( 2{{\sigma }_{{{\mathbf{a}}_{i}}{{\mathbf{b}}_{i}}}} + {{C}_{2}} \right)}{\left( \mu _{{{\mathbf{a}}_{i}}}^{2} + \mu _{{{\mathbf{b}}_{i}}}^{2} + {{C}_{1}} \right) \left( \sigma _{{{\mathbf{a}}_{i}}}^{2} + \sigma _{{{\mathbf{b}}_{i}}}^{2} + {{C}_{1}} \right)}} $},
\end{equation}%
where $\mathbf{A}$ and $\mathbf{B}$ represent fully sampled and reconstructed SSOS images, respectively; ${{\mu }_{{{\mathbf{a}}_{i}}}}$ , ${{\mu }_{{{\mathbf{b}}_{i}}}}$, ${{\sigma }_{{{\mathbf{a}}_{i}}}}$, ${{\sigma }_{{{\mathbf{b}}_{i}}}}$ and ${{\sigma }_{{{\mathbf{a}}_{i}}{{\mathbf{b}}_{i}}}}$ respectively denote the means, standard deviations and covariance of the local window ${{\mathbf{a}}_{i}}$ and ${{\mathbf{b}}_{i}}$; $\mathbf{M}$ the number of local windows. Constants ${{C}_{1}}$ and ${{C}_{2}}$ are introduced to avoid the case when the denominator multiplied by $\mu _{{{\mathbf{a}}_{i}}}^{2}+\mu _{{{\mathbf{b}}_{i}}}^{2}$ and $\sigma _{{{\mathbf{a}}_{i}}}^{2}+\sigma _{{{\mathbf{b}}_{i}}}^{2}$ is close to zero. A higher MSSIM indicates higher detail preservation in reconstruction.

\subsection{Experiments on Phantom Data}
The fully sampled Phantom data (Fig. \ref{fig_results_phantom} (a)) were acquired from a 3T SIEMENS Trio whole-body scanner (Siemens Healthcare, Erlangen, Germany) equipped with a 32-coil head coil, using 2D T2-weighted turbo spin echo sequence (matrix size = $384\times384$, TR/TE = 2000 ms /9.7 ms, FOV = 230 mm$\times$187 mm, slice thickness = 5 mm). The acquired data of 32 coils were compressed into eight virtual coils \cite{2013_ESPIRIT}.
Reconstructions of phantom data are depicted in Fig. \ref{fig_results_phantom}. GRAPPA (Fig. \ref{fig_results_phantom} (b)) and ALOHA (Fig. \ref{fig_results_phantom} (d)) reconstructed images exhibit obvious ringing artifacts in both the white background region and gray objects. $\ell_1$-SPIRiT (Fig. \ref{fig_results_phantom} (c)) produces good reconstructed image with better artifacts suppression. But, with closer inspection, visible noise of the left-bottom oblique sectors can be found. In comparison, STDLR-SPIRiT preserves image resolution and sharped edges (Fig. \ref{fig_results_phantom} (e)). Also, the difference images and quality metric in Table \ref{Table_3} serve to demonstrate STDLR-SPIRiT possessing the lowest reconstruction error (Fig. \ref{fig_results_phantom} (j)) and highest MMSIM.

\begin{table}[htbp]
  \footnotesize
  \centering
  \caption{RLNE/MSSIM FOR PHANTOM AND HUMAN BRAIN DATA RECONSTRUCTIONS.}
  \label{Table_3}
    \begin{tabular}{m{0.6cm}<{\centering}m{1.2cm}<{\centering}m{1.3cm}<{\centering}m{1cm}<{\centering}m{2cm}<{\centering}}
    \toprule
    \multicolumn{1}{c}{Images} & GRAPPA & $\ell_1$-SPIRiT & ALOHA & STDLR-SPIRiT \\
    \midrule
    \multicolumn{1}{c}{\multirow{2}[2]{*}{Fig. 6}}  & \multicolumn{1}{c}{0.0862} & \multicolumn{1}{c}{0.0692} & \multicolumn{1}{c}{0.0827} & \multicolumn{1}{c}{\textbf{0.0470}} \\
          & /0.8951 & /0.9674 & /0.9512 & \textbf{/0.9831} \\
    \midrule
    \multicolumn{1}{c}{\multirow{2}[2]{*}{Fig. 7}}  & \multicolumn{1}{c}{0.1335} & \multicolumn{1}{c}{0.0866} & \multicolumn{1}{c}{0.1117} & \multicolumn{1}{c}{\textbf{0.0735}} \\
          & /0.9609 & /0.9868 & /0.9791 & \textbf{/0.9919} \\
    \midrule
    \multicolumn{1}{c}{\multirow{2}[2]{*}{Fig. 8}}  & \multicolumn{1}{c}{0.2726} & \multicolumn{1}{c}{0.1891} & \multicolumn{1}{c}{0.1840} & \multicolumn{1}{c}{\textbf{0.1187}} \\
          & /0.9067 & /0.9485 & /0.9523 & \textbf{/0.9777} \\
    \midrule
    \multicolumn{1}{c}{\multirow{2}[2]{*}{Fig. 9}} & \multicolumn{1}{c}{0.1127} & \multicolumn{1}{c}{0.0848} & \multicolumn{1}{c}{0.0997} & \multicolumn{1}{c}{\textbf{0.0744}} \\
          & /0.9411 & /0.9755 & /0.9723 & \textbf{/0.9833} \\
    \bottomrule
    \end{tabular}%
\end{table}%

\subsection{Experiments on Brain Imaging Data}
Three brain datasets are used in experiments here. The first brain dataset shown in Fig. \ref{fig_results_brain_typical} and Fig. \ref{fig_results_brainData} (a) were two slices acquired from a healthy volunteer using the 2D $T_2$-weighted turbo spin echo sequence (matrix size = 256 $\times$ 256, TR/TE=6100 ms/99 ms, field of view = 220 mm $\times$ 220 mm, slice thickness = 3mm). They were obtained from a 3T SIEMENS Trio whole-body scanner (Siemens Healthcare, Erlangen, Germany) equipped with a 32-coil head coil. Four virtual coils were compressed from the acquired data of 32 coils \cite{2013_ESPIRIT}. The second brain dataset shown in Fig. \ref{fig_results_brain_typical_2} and Fig. \ref{fig_results_brainData} (b) were two slices acquired from a healthy volunteer on a 1.5T Philips MRI scanner (Philips Healthcare Best, the Netherlands) equipped with an 8-coil head coil using the 2D $T_1$-weighted fast-field-echo sequence (matrix size = 256 $\times$ 256, TR/TE = 1700 ms/390 ms, FOV = 230 mm$\times$230 mm, slice thickness = 5 mm). The third brain dataset depicted in Fig. \ref{fig_results_brainData} (c) were acquired from a 3T GE MRI scanner (GE Healthcare, Waukesha, Wis, USA) equipped with a 12-coil head coil using the 2D $T_1$-weighted SPGR sequences (matrix size = 256 $\times$ 256, TR/TE = 400 ms/9 ms, FOV = 240 mm $\times$ 240 mm, slice thickness = 6 mm.).

Serious ringing artifacts remain in the reconstructions by GRAPPA (Fig. \ref{fig_results_brain_typical} (b)) and ALOHA (Fig. \ref{fig_results_brain_typical} (d)). The $\ell_1$-SPIRiT produces the visually similar reconstructed image (Fig. \ref{fig_results_brain_typical} (c)) as STDLR-SPIRiT does. Both methods show great capability of artifacts suppression and present reliable reconstructions.
For better visualization of their differences, we offer zoom-in images of reconstructions. From the zoom-in images, we could observe relatively stronger noise in $\ell_1$-SPIRiT reconstruct image and to some extend, blur of details.
Nevertheless STDLR-SPIRiT restores image with better signal to noise ratio and fine details preservation (Fig. \ref{fig_results_brain_typical} (e)). In addition, the error images illustrate that the proposed STDLR-SPIRiT achieves the lowest reconstruction error (Fig. \ref{fig_results_brain_typical} (j)). Objective criteria in Table \ref{Table_3} indicates that the proposed method produces the lowest reconstruction error and the highest structure similarity to the fully sampled image. Evaluation criteria are consistent with visual inspections mentioned above.
\begin{figure}[htbp]
\setlength{\abovecaptionskip}{0.cm}
\setlength{\belowcaptionskip}{-0.cm}
\centering
\includegraphics[width=3.4in]{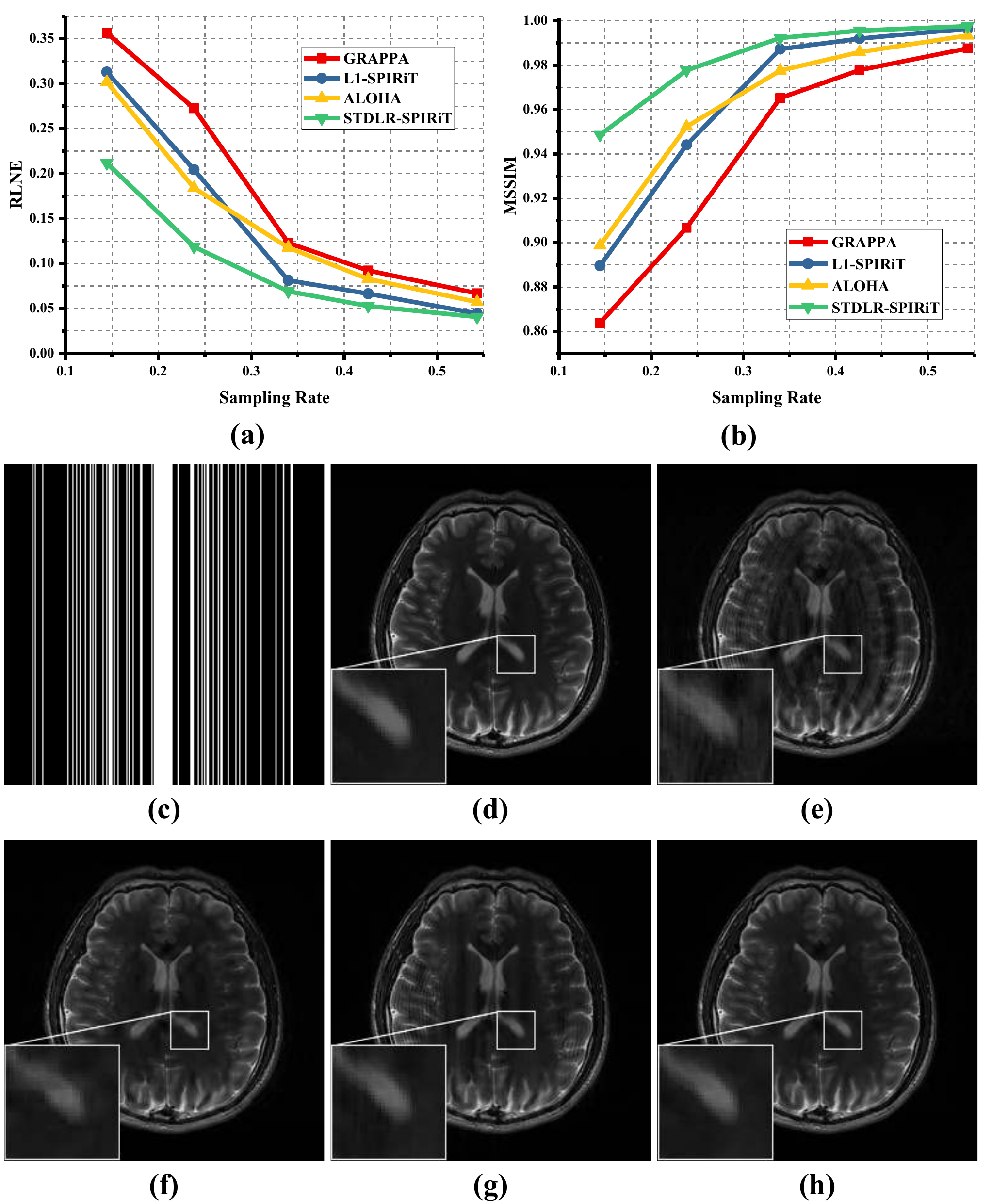}
\caption{The reconstruction results of T2-weighted brain image versus different sampling rates under Cartesian undersampling pattern. (a) and (b) display the variation trends of RLNE and MSSIM respectively; (c) the Cartesian undersampling pattern with a sampling rate of 0.24; (d) an SSOS image of fully sampled data; (e-h) SSOS images of reconstructed results by GRAPPA, $\ell_1$-SPIRiT, ALOHA and STDLR-SPIRiT. Note: The fully sampled images in Figs. \ref{fig_results_brain_typical} and \ref{fig_results_diff_patterns_RLNE} are the same.}
\label{fig_results_diff_patterns_RLNE}
\end{figure}
\begin{figure}[htbp]
\setlength{\abovecaptionskip}{0.cm}
\setlength{\belowcaptionskip}{-0.cm}
\centering
\includegraphics[width=3.4in]{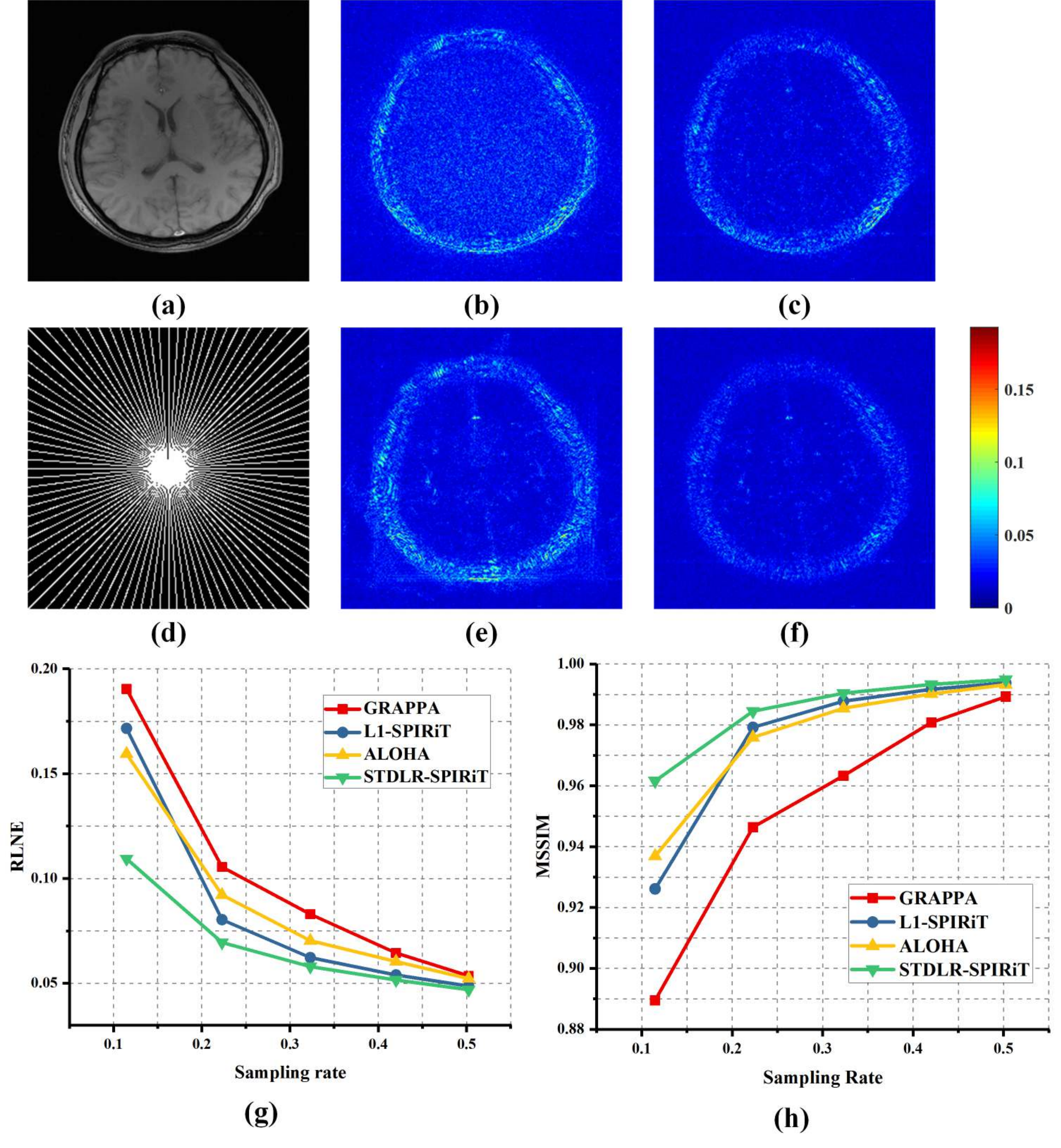}
\caption{Reconstructed brain images using pseudo radial pattern with sampling rate of 0.20. (a) An SSOS image of fully sampled data; (b, c, e, f) reconstruction errors introduced by GRAPPA, $\ell_1$-SPIRiT, ALOHA and STDLR-SPIRiT; (d) pseudo radial undersampling pattern; (g) and (h) display the  RLNE and MSSIM versus sampling rate, respectively.}
\label{fig_results_brain_typical_2}
\end{figure}

As shown in Figs. \ref{fig_results_diff_patterns_RLNE} (a) and (b), with the increase of sampling rates, the growth of MSSIM and decline of RLNE are observed in all methods. Moreover, the STDLR-SPIRiT outperforms other methods under all sampling rates in terms of RLNE and SSIM. Particularly, even at a relatively low sampling rate (0.24), the proposed method still provides reliable reconstructions (Fig. \ref{fig_results_diff_patterns_RLNE} (h)) whereas GRRAPA and ALOHA images carry distinct artifacts. $\ell_1$-SPIRiT produces an image with nice artifacts suppression (Fig. \ref{fig_results_diff_patterns_RLNE} (f)), nonetheless the image is blurred (the zoom-in image) and bear incorrect contrast reconstruction of the withe matter in some areas, for instance, the mottled white matters around the image center. By contrast, STDLR-SPIRiT retains promising image resolution and fine details as the fully sampled image does (Fig. \ref{fig_results_diff_patterns_RLNE} (h)).

\begin{figure}[htbp]
\setlength{\abovecaptionskip}{0.cm}
\setlength{\belowcaptionskip}{-0.cm}
\centering
\includegraphics[width=3.4in]{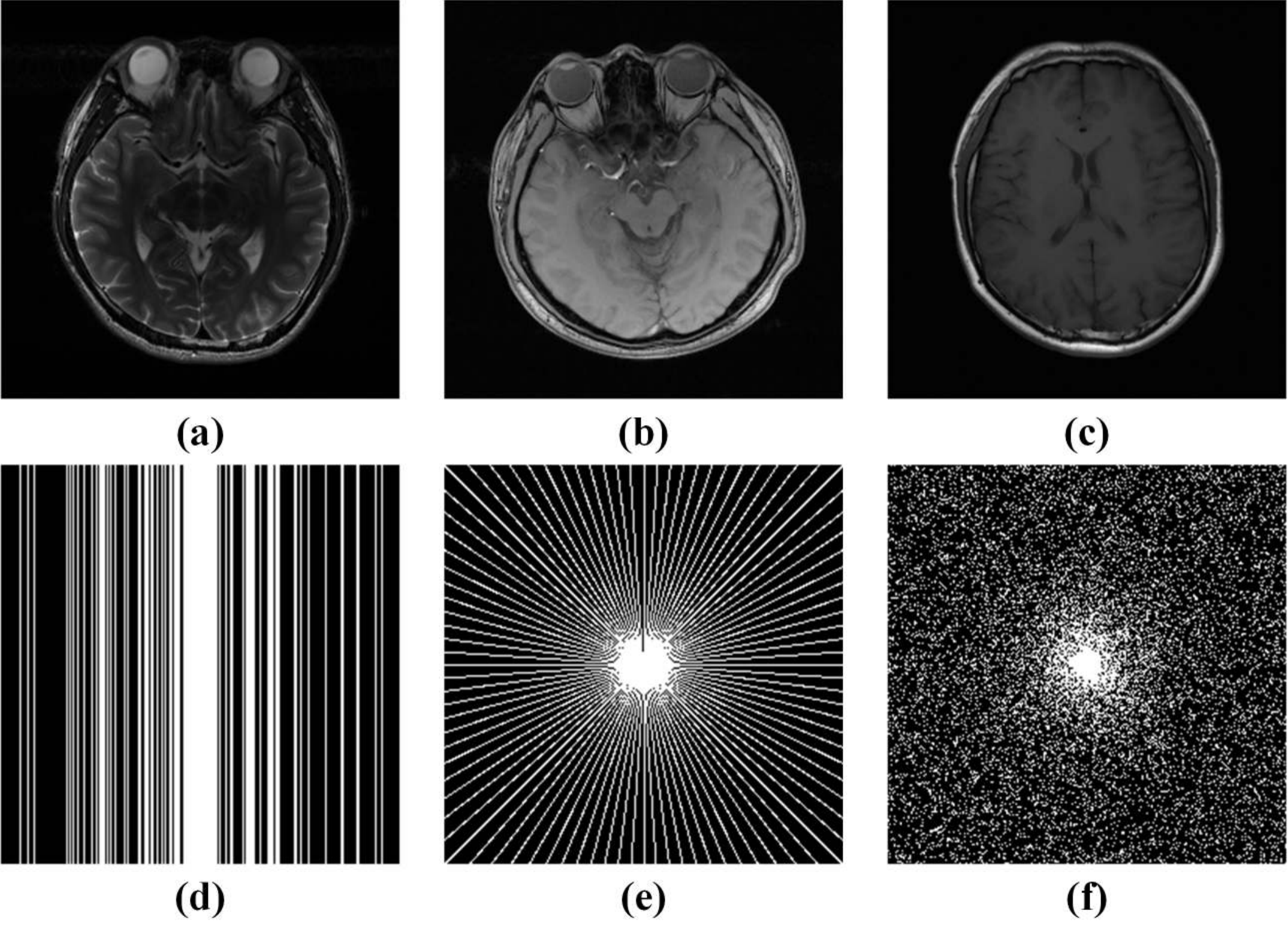}
\caption{More brain images and undersampling patterns. (a-c) Three different brain images; (d) the Cartesian sampling pattern of sampling rate 0.34; (e) the pseudo radial sampling pattern of sampling rate 0.20; (f) the 2D random sampling pattern of sampling rate 0.18.}
\label{fig_results_brainData}
\end{figure}
An experiment with the pseudo radial sampling pattern was conducted, and results were shown in Fig. \ref{fig_results_brain_typical_2}. All tested approaches provide good reconstructions with relatively small error. The overall error of GRAPPA reconstruction seems to be the biggest one, but its error inside the skull has nice random distribution, which would be preferred in applications. $\ell_1$-SPIRiT, ALOHA and STDLR-SPIRiT errors have reached the promising low level, assuring reliable reconstructions. It is worthy of mentioning that STDLR-SPIRiT outperforms compared methods interns of RLNE and MSSIM, and obtains the smallest reconstruction error, which serves as another demonstration for the state-of-the-art performance of the proposed method. Other undersampling patterns and brain images in Fig. \ref{fig_results_brainData} are also used for reconstruction for further validation and the RLNE and MSSIM metrics are summarized in Table \ref{Table 4}. Superior metrics provided by the proposed method imply better reconstructed images.

\begin{table}[htbp]
\footnotesize
\begin{center}
  \centering
  \caption{RLNE/MSSIM Results for Brains in Fig. \ref{fig_results_brainData} Using Different Samplings.}
  \label{Table 4}
    \begin{tabular}{m{0.8cm}<{\centering}m{0.8cm}<{\centering}m{0.8cm}<{\centering}m{1.3cm}<{\centering}m{0.8cm}<{\centering}m{1.0cm}<{\centering}}
    \toprule
    \multicolumn{1}{c}{Images} & \multicolumn{1}{c}{Pattern} & GRAPPA & $\ell_1$-SPIRiT & ALOHA & STDLR-SPIRiT \\
    \midrule
    \multicolumn{1}{c}{\multirow{6}[6]{*}{Fig.10(a)}} & \multicolumn{1}{c}{\multirow{2}[2]{*}{Fig. 10(d)}} & \multicolumn{1}{c}{0.1399} & \multicolumn{1}{c}{0.1110} & \multicolumn{1}{c}{0.1232} & \multicolumn{1}{c}{\textbf{0.0904}} \\
          &       & /0.9626 & /0.9799 & /0.9754 & \textbf{/0.9883} \\
\cmidrule{2-6}          & \multicolumn{1}{c}{\multirow{2}[2]{*}{Fig. 10(e)}} & \multicolumn{1}{c}{0.1222} & \multicolumn{1}{c}{0.1013} & \multicolumn{1}{c}{0.1370} & \multicolumn{1}{c}{\textbf{0.0879}} \\
          &       & /0.9704 & /0.9798 & /0.9722 & \textbf{/0.9869} \\
\cmidrule{2-6}          & \multicolumn{1}{c}{\multirow{2}[2]{*}{Fig. 10(f)}} & \multicolumn{1}{c}{0.1540} & \multicolumn{1}{c}{0.1013} & \multicolumn{1}{c}{0.1103} & \multicolumn{1}{c}{\textbf{0.0832}} \\
          &       & /0.9523 & /0.9788 & /0.9764 & \textbf{/0.9878} \\
    \midrule
    \multicolumn{1}{c}{\multirow{6}[6]{*}{Fig.10(b)}} & \multicolumn{1}{c}{\multirow{2}[2]{*}{Fig. 10(d)}} & \multicolumn{1}{c}{0.1854} & \multicolumn{1}{c}{0.1115} & \multicolumn{1}{c}{0.1132} & \multicolumn{1}{c}{\textbf{0.0903}} \\
          &       & /0.8665 & /0.9518 & /0.9490 & \textbf{/0.9682} \\
\cmidrule{2-6}          & \multicolumn{1}{c}{\multirow{2}[2]{*}{Fig. 10(e)}} & \multicolumn{1}{c}{0.1260} & \multicolumn{1}{c}{0.1005} & \multicolumn{1}{c}{0.1103} & \multicolumn{1}{c}{\textbf{0.0846}} \\
          &       & /0.9259 & /0.9632 & /0.9617 & \textbf{/0.9754} \\
\cmidrule{2-6}          & \multicolumn{1}{c}{\multirow{2}[2]{*}{Fig. 10(f)}} & \multicolumn{1}{c}{0.1706} & \multicolumn{1}{c}{0.1018} & \multicolumn{1}{c}{0.1011} & \multicolumn{1}{c}{\textbf{0.0851}} \\
          &       & /0.8693 & /0.9574 & /0.9584 & \textbf{/0.9725} \\
    \midrule
    \multicolumn{1}{c}{\multirow{6}[6]{*}{Fig.10(c)}} & \multicolumn{1}{c}{\multirow{2}[2]{*}{Fig. 10(d)}} & \multicolumn{1}{c}{0.1306} & \multicolumn{1}{c}{0.0830} & \multicolumn{1}{c}{0.0916} & \multicolumn{1}{c}{\textbf{0.0779}} \\
          &       & /0.9362 & /0.9883 & /0.9830 & \textbf{/0.9902} \\
\cmidrule{2-6}          & \multicolumn{1}{c}{\multirow{2}[2]{*}{Fig. 10(e)}} & \multicolumn{1}{c}{0.1161} & \multicolumn{1}{c}{0.0864} & \multicolumn{1}{c}{0.1073} & \multicolumn{1}{c}{\textbf{0.0822}} \\
          &       & /0.9624 & /0.9880 & /0.9827 & \textbf{/0.9905} \\
\cmidrule{2-6}          & \multicolumn{1}{c}{\multirow{2}[2]{*}{Fig. 10(f)}} & \multicolumn{1}{c}{0.1269} & \multicolumn{1}{c}{0.0859} & \multicolumn{1}{c}{0.0952} & \multicolumn{1}{c}{\textbf{0.0807}} \\
          &       & /0.9437 & /0.9876 & /0.9851 & \textbf{/0.9908} \\
    \bottomrule
    \end{tabular}%
    \end{center}
\end{table}%

\section{Discussions}\label{Section:discussions}
\subsection{Discussion on ACS}
\begin{figure}[htbp]
\setlength{\abovecaptionskip}{0.cm}
\setlength{\belowcaptionskip}{-0.cm}
\centering
\includegraphics[width=3.4in]{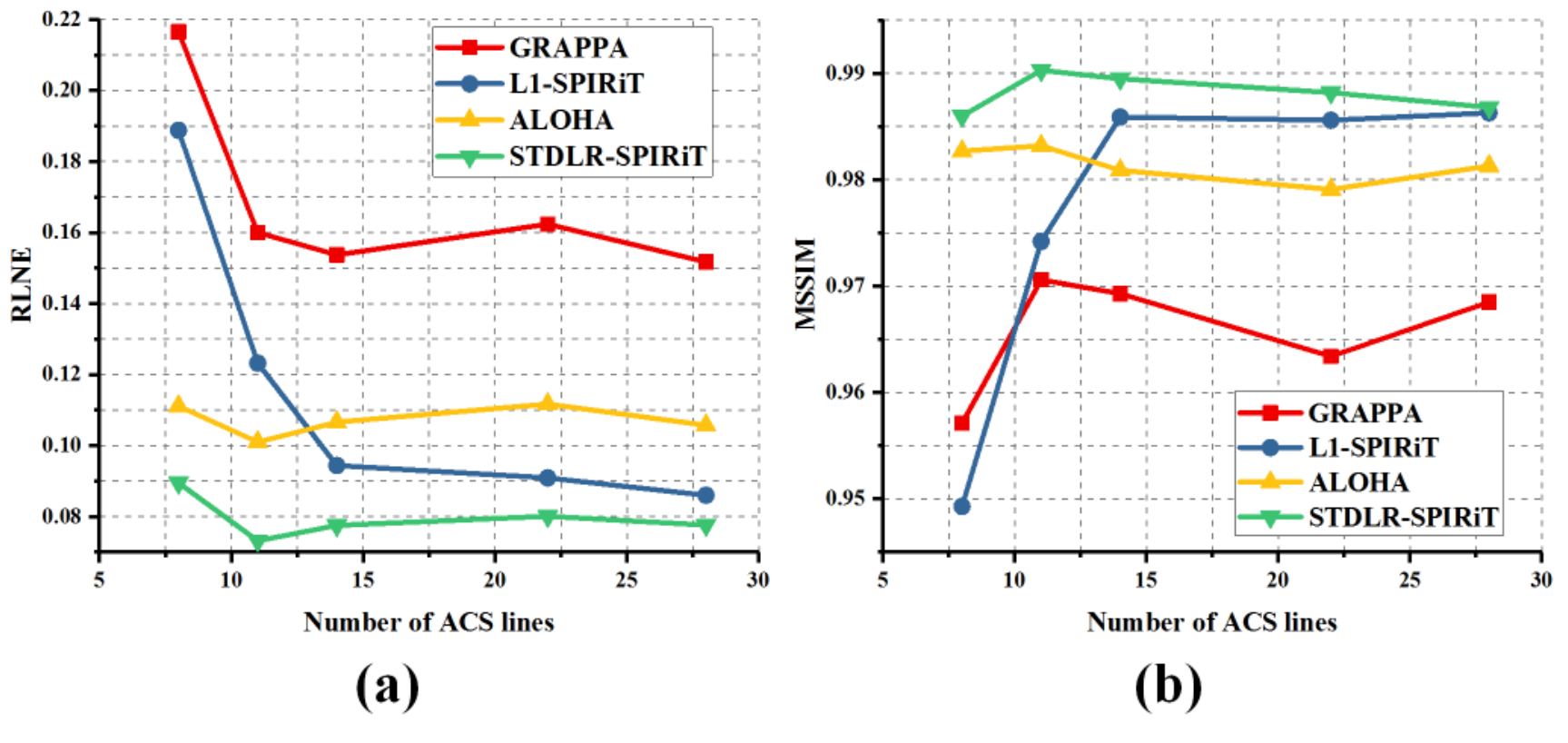}
\caption{The variation trends of RLNE (a) and MSSIM (b) versus the number of ACS lines. This experiment is carried out with a series of Gaussian Cartesian sampling patterns but all at the same sampling rate of 0.34. The difference between each sampling pattern is the number of ACS lines. Note: all experiments are conducted on the brain data shown in Fig. \ref{fig_results_brain_typical} (a).}
\label{fig_results_ACS_RLNE}
\end{figure}

The proposed STDLR-SPIRiT is robust to the number of ACS lines. As shown in Fig. \ref{fig_results_ACS_RLNE} (a), the reconstruction errors - RLNE - of STDLR-SPIRiT the smallest among all methods. The SPIRiT monotonically reduces reconstruction errors when the number of ACS lines increases. When ACS signals ($>14$ lines) are enough, the $\ell_1$-SPIRiT achieves much lower error than GRAPPA and ALOHA. These observations indicate that the SPIRiT effectively mines the self-consistency of k-space via ACS, resulting in improved image reconstruction. However, with a decrease of ACS lines ($<14$ lines), the accuracy of kernels estimated in the $\ell_1$-SPIRiT would decrease, leading to a higher RLNE than that of ALOHA. This observation suggests that low-rankness in ALOHA is a very valuable property to regularize image reconstructions and may alleviate the influence of the decline of ACS numbers. The proposed STDLR-SPIRiT inherits advantages of SPIRiT and ALOHA, leading to the lowest reconstruction error under all the number of ACS lines.

\subsection{Discussion on Parameter Settings}
In this subsection, the effect of parameters setting in STDLR-SPIRiT will be discussed, including regularization parameters ${{\lambda }_{1}}$ and ${{\lambda }_{2}}$ as well as pencil parameters ${{k}_{1}}$ and ${{k}_{2}}$.
The brain image in Fig. \ref{fig_results_brain_typical} (a) and the sampling pattern in Fig. \ref{fig_results_brain_typical} (f) were used to perform reconstructions. Typical settings are ${{\lambda }_{1}}={{10}^{4}}$, ${{\lambda}_{2}}={{10}^{6}}$ and ${{k}_{1}}={{k}_{2}}=23$. When one parameter is analyzed, other parameters are set as the typical values.

\begin{figure}[htbp]
\setlength{\abovecaptionskip}{0.cm}
\setlength{\belowcaptionskip}{-0.cm}
\centering
\includegraphics[width=3.2in]{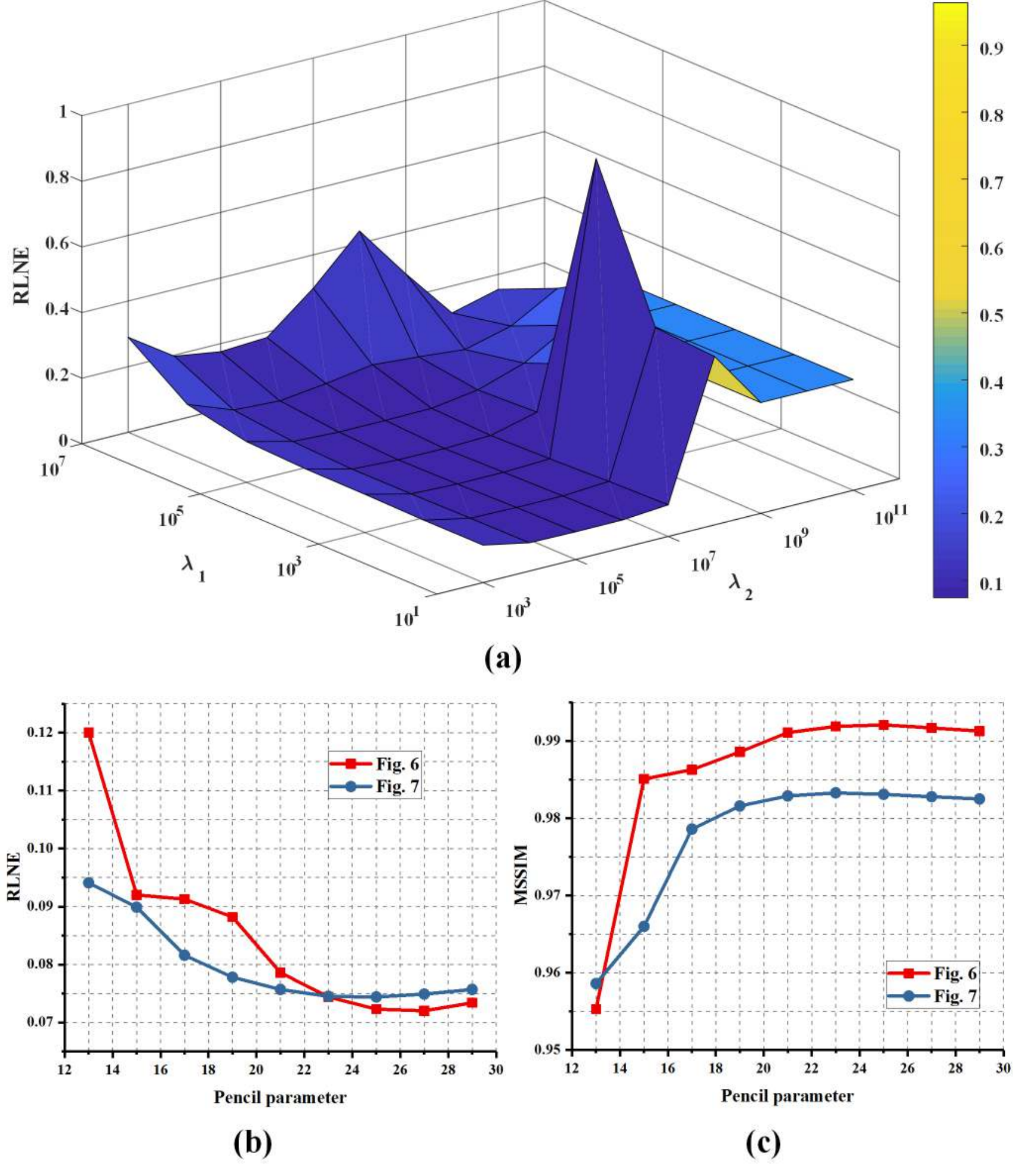}
\caption{Variation trends of reconstruction errors by STDLR-SPIRiT with different parameter settings. (a) The RLNE versus regularization parameters; (b-c) the RLNE and MSSIM versus pencil parameters. Note: all experiments are conducted on the brain data shown in Fig. \ref{fig_results_brain_typical} (a) under Gaussian Cartesian undersampling pattern shown in Fig. \ref{fig_results_brain_typical} (f).}
\label{fig_results_parameter_setting}
\end{figure}

The reconstruction errors versus different values of regularization parameters are depicted in Fig. \ref{fig_results_parameter_setting} (a). The selection of ${{\lambda}_{1}}$ and ${{\lambda}_{2}}$ relies on noise level and the amount of ACS data for kernel estimation. As can be seen in Fig. \ref{fig_results_parameter_setting} (a), under fixed noise level and sampling pattern, there exists a wide range of ${{\lambda}_{1}}$ (${{10}^{1}}\le {{\lambda}_{1}}\le {{10}^{5}}$) and ${{\lambda}_{2}}$ (${{10}^{4}}\le {{\lambda}_{2}}\le {{10}^{7}}$) leading to relatively low reconstruction errors. Too small or too large values of ${{\lambda}_{1}}$ and ${{\lambda}_{2}}$ produce higher reconstruction errors. When the amount of data for SPIRiT kernel estimation are not adequate, we suggest choosing a relatively small value for ${{\lambda}_{1}}$ to alleviate the effect of less accurate kernel estimation. In this paper, we set ${{\lambda }_{1}}={{10}^{4}}$ and ${{\lambda}_{2}}={{10}^{6}}$ for all experiments.

The values of ${{k}_{1}}$ and ${{k}_{2}}$ determine the size of the block Hankel matrix, which in turn affect the rank of the matrix. The rank of block Hankel matrix is related to the sparsity level of transform domain. When ${{k}_{1}}$ and ${{k}_{2}}$ are too small (for example, less than $13$ here), the rank of the block Hankel matrix is smaller than the sparsity level, leading to larger reconstruction errors, as shown in Figs. \ref{fig_results_parameter_setting} (b) and (c). When ${{k}_{1}}$ and ${{k}_{2}}$ are bigger than the sparsity level of the transform domain, the reconstruction errors are not sensitive to ${{k}_{1}}$ and ${{k}_{2}}$ values. However, big values of ${{k}_{1}}$ and ${{k}_{2}}$ cost large computation load. Therefore, there is a tradeoff between reconstruction quality and computation load, we empirically set ${{k}_{1}}$ and ${{k}_{2}}$ in the range from 21 to 25.

\subsection{Algorithm Convergence}
\begin{figure}[htbp]
\setlength{\abovecaptionskip}{0.cm}
\setlength{\belowcaptionskip}{-0.cm}
\centering
\includegraphics[width=3.4in]{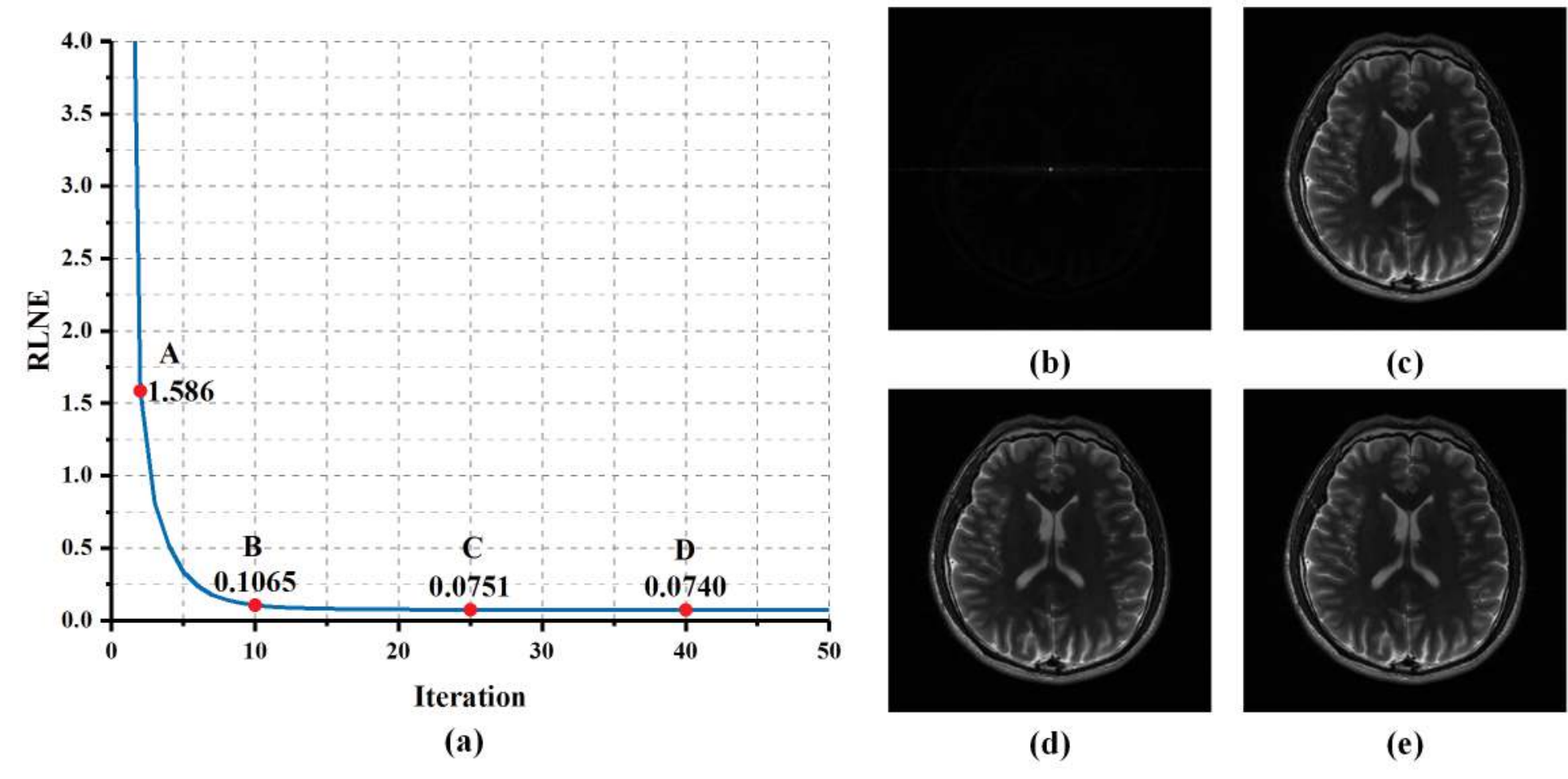}
\caption{Empirical convergence results. (a) is the empirical convergence of STDLR-SPIRiT; (b-e) are reconstructions of STDLR-SPIRiT at A to D iterations depicted in (a), respectively.}
\label{fig_convergence}
\end{figure}

We adopted ADMM \cite{2011_ADMM} to solve the non-convex STDLR-SPIRiT model in \ref{(16)}.  It has been shown that ADMM can be used for successfully solving non-convex optimization problems \cite{2016_ALOHA,2018_TBME_Hengfa,2017_ACCESS_Guo}. Empirically, we observed that RLNE decreases to a relatively small value as iteration increases (e.g. the point C depicted in \ref{fig_convergence}(a)), and then stabilizes in a small RLNE level. For example, as the iteration rises (from point A to C), proposed approach achieving increasingly better reconstruction, and eventual provides reliable reconstruction (\ref{fig_convergence}(c-d)).

\subsection{Comparison with Other State-of-the-art Methods}
In this sub-section, we compare the performance of the proposed method with other state-of-the-art low-rank structured matrix approaches AC-LORAKS(S) \cite{2015_AC_LORAKS} and GSLR \cite{2018_Hu_LR_TMI} on two data set. 

\begin{figure}[htbp]
\setlength{\abovecaptionskip}{0.cm}
\setlength{\belowcaptionskip}{-0.cm}
\centering
\includegraphics[width=3.4in]{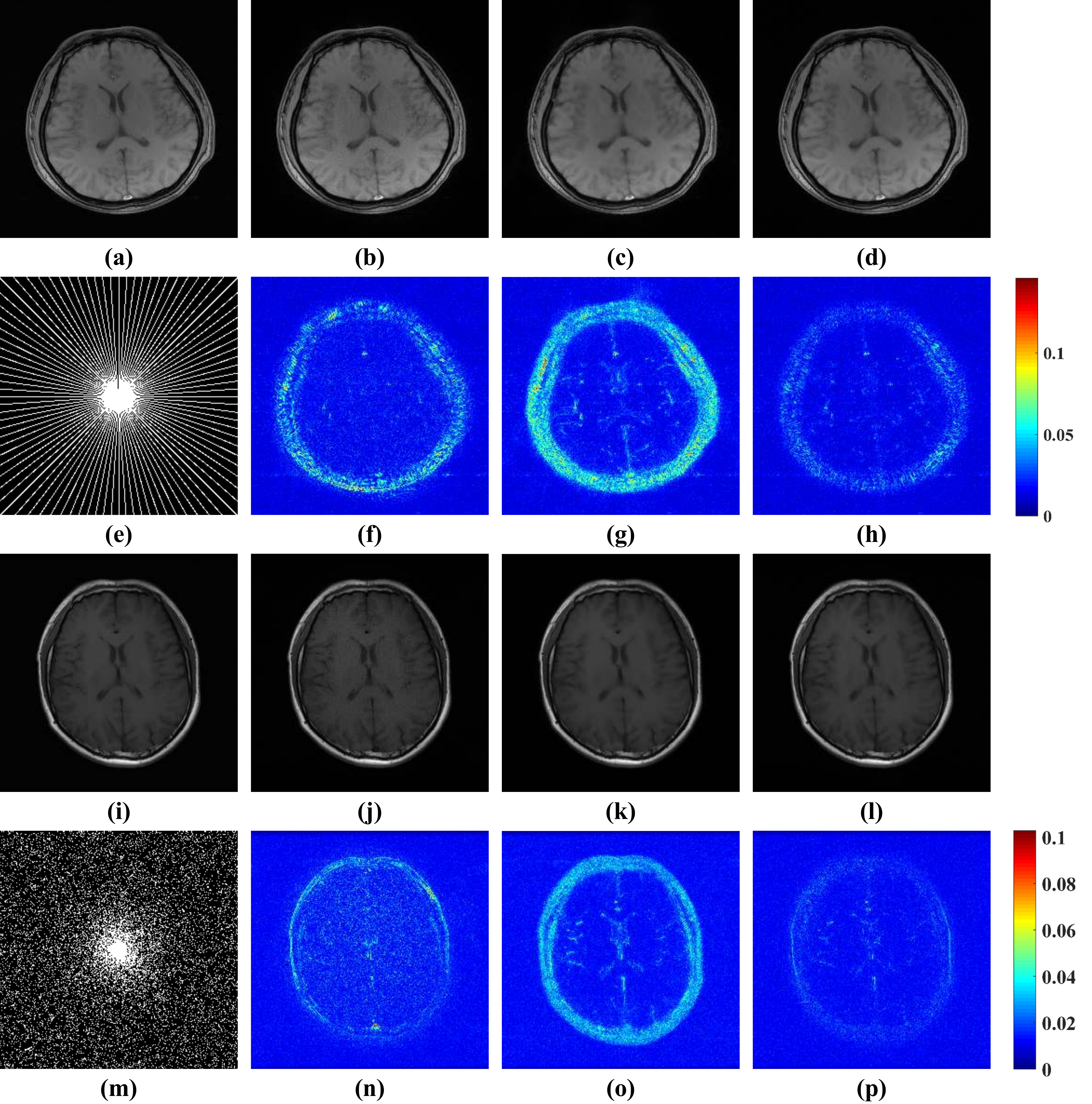}
\caption{Reconstructions of multi-coils brain images. (a) An SSOS image of fully sampled image; (b-d) SSOS images of reconstructed images by AC-LORAKS(S), GSLR and STDLR-SPIRiT; (e) the pseudo radial undersampling pattern with a sampling rate of 0.20; (f-h) the reconstruction error distribution corresponding to the above methods; (i) an SSOS image of the fully sampled image; (j-l) SSOS images of reconstructed images by AC-LORAKS(S), GSLR and STDLR-SPIRiT; (m) the 2D random undersampling pattern with a sampling rate 0.18; (n-p) the reconstruction error distribution corresponding to the above methods. Note: The first tested image and the corresponding pseudo pattern were from Fig. \ref{fig_results_brain_typical_2}. The second tested image and the corresponding 2D random pattern were from Fig. \ref{fig_results_brainData}.}
\label{fig_results_otherMethods}
\end{figure}
Here we choose AC-LORAKS(S) instead of other LORAKS variants since AC-LORAKS and the proposed method both exploit the calibration data, but, please note AC-LORAKS uses it for only algorithm acceleration. The letter 'S' in the bracket of AC-LORAKS(S) denotes the applying of phase constraint in AC-LORAKS. It has been shown that LORAKS with phase constraint produces the best result compared with other constraints \cite{2014_LORAKS}, such as, support constraint. Regarding of GSLR, it cascades the weighting matrices on the horizontal and vertical directions into a fatty low-rank matrix, which increases the computational complexity, and suggests that imposing simultaneous weighting on multi-coils images would be difficult because the resulting matrix would be of vast dimension. Note that the GSLR code shared by the author reconstructs multi-coils images channel-by-channel, that is to say, GSLR makes no use of the correlations among coils. This indicates GSLR reconstruction may be inferior to that of STDLR-SPIRiT.

Reconstructions of the tested images were presented in Fig. \ref{fig_results_otherMethods}, in which AC-LORAKS(S), GSLR and STDLR-SPIRiT all provide images with nice artifacts suppression. However, apparent noise could be inspected in AC-LORAKS(S) reconstruction (Figs. \ref{fig_results_otherMethods}(b) and (j)), but this may not be a bad thing because it allows for a better visualization of the image. In addition, blurring of image can be observed in GSLR reconstruction (Figs. \ref{fig_results_otherMethods}(g) and (k)). Whilst STDLR-SPIRiT produces promising reconstruction with satisfying noise level and resolution (Figs. \ref{fig_results_otherMethods}(d) and (i)). The error images shown in Figs. \ref{fig_results_otherMethods} (f-h) and (n-p) reveals STDLR-SPIRiT embracing the lowest structure error compared with AC-LORAKS(S) and GSLR. The GSLR yielding obvious error compared with the proposed method verifies the aforementioned discussion about GSLR.

\section{Conclusion}\label{Section:conclusion}
A simultaneous two-directional low-rank structured Hankel matrix reconstruction approach, STDLR-SPIRiT, is proposed for accelerated data acquisition in parallel MRI. The proposed approach simultaneously utilizes the self-consistency of k-space data and the low-rankness of the weighted k-space data, which yields a better exploitation of correlation among inter- and intra-coils. The STDLR-SPIRiT provides robust reconstruction with a limited number of auto-calibration signals. Experiments on phantom and brain MRI demonstrate that the proposed approach achieves superior performance in artifacts suppression and edge preservation than the state-of-the-art methods. Although a singular value decomposition-free algorithm is derived to reduce the computation load, faster algorithms are still expected in the future \cite{2014_LORAKS,2016_Off_the_grid}. Besides, the low rank Hankel matrix is highly related to exponential functions, and latest reconstruction methods of multi-dimensional exponentials \cite{2018_Vandermonde_Jiaxi, 2017_TSP_Jiaxi} may improve the Hankel-based MRI reconstruction or be extended into higher dimensional MRI in fast imaging.

\section*{Acknowledgments}
Xiaobo Qu is grateful to Prof. Chun Yuan for hosting his visit at University of Washington. This work was supported in part by National Key R\&D Program of China (2017YFC0108703), National Natural Science Foundation of China (61571380, 61871341, 61811530021, U1632274, 61672335 and 61671399), Natural Science Foundation of Fujian Province of China (2018J06018), Fundamental Research Funds for the Central Universities (20720180056), Science and Technology Program of Xiamen (3502Z20183053), and China Scholarship Council (201806315010).

\begin{table}[htbp]
  \footnotesize
  \centering
  \caption{PARAMETERS FOR GRAPPA.}
    \begin{tabular}{m{0.8cm}<{\centering}m{1.4cm}<{\centering}m{1.3cm}<{\centering}m{2.3cm}<{\centering}}
    \toprule
    \multicolumn{1}{c}{Images} & Pattern & Kernel size & Kernel calibration parameters \\
    \midrule
    \multicolumn{1}{c}{Fig. 6} & /     & $9 \times 9$   & 0.03 \\
    \midrule
    \multicolumn{1}{c}{Fig. 7} & /     & $9 \times 9$   & 0.05 \\
    \midrule
    \multicolumn{1}{c}{Fig. 8} & /     & $9 \times 9$   & 0.05 \\
    \midrule
    \multicolumn{1}{c}{Fig. 9} & /     & $5 \times 5$   & 0.10 \\
    \midrule
    \multicolumn{1}{c}{\multirow{3}[6]{*}{Fig. 10 (a)}} & Fig. 10 (d) & $9 \times 9$   & 0.05 \\
    \cmidrule{2-4}          & Fig. 10 (e) & $9 \times 9$   & 0.10 \\
    \cmidrule{2-4}          & Fig. 10 (f) & $7 \times 7$   & 0.10 \\
    \midrule
    \multicolumn{1}{c}{\multirow{3}[6]{*}{Fig. 10 (b)}} & Fig. 10 (d) & $5 \times 5$   & 0.03 \\
    \cmidrule{2-4}          & Fig. 10 (e) & $5 \times 5$   & 0.10 \\
    \cmidrule{2-4}          & Fig. 10 (f) & $5 \times 5$   & 0.03 \\
    \midrule
    \multicolumn{1}{c}{\multirow{3}[6]{*}{Fig. 10 (c)}} & Fig. 10 (d) & $9 \times 9$   & 0.10 \\
    \cmidrule{2-4}          & Fig. 10 (e) & $5 \times 5$   & 0.10 \\
    \cmidrule{2-4}          & Fig. 10 (f) & $5 \times 5$   & 0.10 \\
    \bottomrule
    \end{tabular}%
  \label{Table 5}%
\end{table}%

\begin{table}[htbp]
  \footnotesize
  \centering
  \caption{PARAMETERS FOR $\ell_1$-SPIRIT.}
    \begin{tabular}{m{0.8cm}<{\centering}m{1.4cm}<{\centering}m{0.8cm}<{\centering}m{1.5cm}<{\centering}m{1.5cm}<{\centering}}
    \toprule
    \multicolumn{1}{c}{Images} & Sampling pattern & Kernel size & Kernel calibration parameter & Wavelet regularization parameter \\
    \midrule
    \multicolumn{1}{c}{Fig. 6} & /     & $9 \times 9$   & 0.0010 & 0.0010 \\
    \midrule
    \multicolumn{1}{c}{Fig. 7} & /     & $7 \times 7$   & 0.0030 & 0.0015 \\
    \midrule
    \multicolumn{1}{c}{Fig. 8} & /     & $7 \times 7$   & 0.0030 & 0.0015 \\
    \midrule
    \multicolumn{1}{c}{Fig. 9} & /     & $5 \times 5$   & 0.0030 & 0.0005 \\
    \midrule
    \multicolumn{1}{c}{\multirow{3}[6]{*}{Fig. 10 (a)}} & Fig. 10 (d) & $5 \times 5$   & 0.0050 & 0.0015 \\
    \cmidrule{2-5}          & Fig. 10 (e) & $7 \times 7$   & 0.0050 & 0.0015 \\
    \cmidrule{2-5}          & Fig. 10 (f) & $5 \times 5$   & 0.0010 & 0.0015 \\
    \midrule
    \multicolumn{1}{c}{\multirow{3}[6]{*}{Fig. 10 (b)}} & Fig. 10 (d) & $7 \times 7$   & 0.0010 & 0.0015 \\
    \cmidrule{2-5}          & Fig. 10 (e) & $5 \times 5$   & 0.0050 & 0.0015 \\
    \cmidrule{2-5}          & Fig. 10 (f) & $5 \times 5$   & 0.0010 & 0.0015 \\
    \midrule
    \multicolumn{1}{c}{\multirow{3}[6]{*}{Fig. 10 (c)}} & Fig. 10 (d) & $5 \times 5$   & 0.0030 & 0.0015 \\
    \cmidrule{2-5}          & Fig. 10 (e) & $7 \times 7$   & 0.0300  & 0.0015 \\
    \cmidrule{2-5}          & Fig. 10 (f) & $5 \times 5$   & 0.0050  & 0.0015 \\
    \bottomrule
    \end{tabular}%
  \label{Table 6}%
\end{table}%

\section*{Supplementary Material}\label{SI}
We ran reconstruction experiments using compared approaches with a series of combinations of model-specific parameters subjecting to each tested data, and the parameters of each method allowing the lowest RLNEs were chosen. The detail parameter settings of GRAPPA and $\ell_1$-SPIRiT are listed in TABLE \ref{Table 5} and TABLE \ref{Table 6}. As for ALOHA, two levels of pyramidal decomposition are adopted with LMaFit tolerances $10^{-1}$, $10^{-2}$. For our proposed method, we set ${k_1} = {k_2} = 23$, ${\lambda _1} = {10^4}$ and ${\lambda _2} = {10^6}$ for all experiments.

\ifCLASSOPTIONcaptionsoff
  \newpage
\fi

\bibliographystyle{IEEEtran}
\bibliography{IEEEabrv,myLib}

\begin{thebibliography}{10}
\providecommand{\url}[1]{#1}
\csname url@samestyle\endcsname
\providecommand{\newblock}{\relax}
\providecommand{\bibinfo}[2]{#2}
\providecommand{\BIBentrySTDinterwordspacing}{\spaceskip=0pt\relax}
\providecommand{\BIBentryALTinterwordstretchfactor}{4}
\providecommand{\BIBentryALTinterwordspacing}{\spaceskip=\fontdimen2\font plus
\BIBentryALTinterwordstretchfactor\fontdimen3\font minus
  \fontdimen4\font\relax}
\providecommand{\BIBforeignlanguage}[2]{{%
\expandafter\ifx\csname l@#1\endcsname\relax
\typeout{** WARNING: IEEEtran.bst: No hyphenation pattern has been}%
\typeout{** loaded for the language `#1'. Using the pattern for}%
\typeout{** the default language instead.}%
\else
\language=\csname l@#1\endcsname
\fi
#2}}
\providecommand{\BIBdecl}{\relax}
\BIBdecl

\bibitem{2017_review_PI}
J.~Hamilton, D.~Franson, and N.~Seiberlich, ``Recent advances in parallel
  imaging for {MRI},'' \emph{Progress in Nuclear Magnetic Resonance
  Spectroscopy}, vol. 101, pp. 71--95, 2017.

\bibitem{1999_SENSE}
K.~P. Pruessmann, M.~Weiger, M.~B. Scheidegger, and P.~Boesiger, ``{SENSE}:
  Sensitivity encoding for fast {MRI},'' \emph{Magnetic Resonance in Medicine},
  vol.~42, pp. 952--962, 1999.

\bibitem{2002_GRAPPA}
M.~A. Griswold, P.~M. Jakob, R.~M. Heidemann, M.~Nittka, V.~Jellus, J.~Wang,
  B.~Kiefer, and A.~Haase, ``Generalized autocalibrating partially parallel
  acquisitions ({GRAPPA}),'' \emph{Magnetic Resonance in Medicine}, vol.~47,
  no.~6, pp. 1202--1210, 2002.

\bibitem{2010_SPIRiT}
M.~Lustig and J.~M. Pauly, ``{SPIRiT}: Iterative self-consistent parallel
  imaging reconstruction from arbitrary \textit{k}-space,'' \emph{Magnetic
  Resonance in Medicine}, vol.~64, no.~2, pp. 457--71, 2010.

\bibitem{2014_MRM_Lustig}
P.~J. Shin, P.~E. Larson, M.~A. Ohliger, M.~Elad, J.~M. Pauly, D.~B. Vigneron,
  and M.~Lustig, ``Calibrationless parallel imaging reconstruction based on
  structured low-rank matrix completion,'' \emph{Magnetic Resonance in
  Medicine}, vol.~72, no.~4, pp. 959--970, 2014.

\bibitem{2007_Sparse_MRI}
J.~M.~P. M.~Lustig, D.~Donoho, ``Sparse {MRI}: {T}he application of compressed
  sensing for rapid {MR} imaging,'' \emph{Magnetic Resonance in Medicine},
  vol.~58, no.~6, pp. 1182--1195, 2007.

\bibitem{2007_MRI_TV}
K.~T. Block, M.~Uecker, and J.~Frahm, ``Undersampled radial {MRI} with multiple
  coils. {I}terative image reconstruction using a total variation constraint,''
  \emph{Magnetic Resonance in Medicine}, vol.~57, no.~6, pp. 1086--98, 2007.

\bibitem{2017_PCS_EMBC}
S.~V. Eslahi, P.~V. Dhulipala, C.~Shi, G.~Xie, and J.~X. Ji, ``Parallel
  compressive sensing in a hybrid space: {A}pplication in interventional
  {MRI},'' in \emph{2017 39th Annual International Conference of the IEEE
  Engineering in Medicine and Biology Society (EMBC)}, 2017, Conference
  Proceedings, pp. 3260--3263.

\bibitem{2016_pFISTA}
Y.~Liu, Z.~Zhan, J.-F. Cai, D.~Guo, Z.~Chen, and X.~Qu, ``Projected iterative
  soft-thresholding algorithm for tight frames in compressed sensing magnetic
  resonance imaging,'' \emph{IEEE Transactions on Medical Imaging}, vol.~35,
  no.~9, pp. 2130--2140, 2016.

\bibitem{2010_xiaobo}
X.~Qu, W.~Zhang, D.~Guo, C.~Cai, S.~Cai, and Z.~Chen, ``Iterative thresholding
  compressed sensing {MRI} based on contourlet transform,'' \emph{Inverse
  Problems in Science and Engineering}, vol.~18, no.~6, pp. 737--758, 2010.

\bibitem{2018_MIA_reweighted_sparse_algorithm}
C.~Chen, L.~He, H.~Li, and J.~Huang, ``Fast iteratively reweighted least
  squares algorithms for analysis-based sparse reconstruction,'' \emph{Medical
  Image Analysis}, vol.~49, pp. 141--152, 2018.

\bibitem{2016_TBME_Zhifang}
Z.~Zhan, J.~Cai, D.~Guo, Y.~Liu, Z.~Chen, and X.~Qu, ``Fast multiclass
  dictionaries learning with geometrical directions in {MRI} reconstruction,''
  \emph{IEEE Transactions on Biomedical Engineering}, vol.~63, no.~9, pp.
  1850--1861, 2016.

\bibitem{2016_MIA_Lai}
Z.~Lai, X.~Qu, Y.~Liu, D.~Guo, J.~Ye, Z.~Zhan, and Z.~Chen, ``Image
  reconstruction of compressed sensing {MRI} using graph-based redundant
  wavelet transform,'' \emph{Medical Image Analysis}, vol.~27, pp. 93--104,
  2016.

\bibitem{2011_Ravishankar}
S.~Ravishankar and Y.~Bresler, ``{MR} image reconstruction from highly
  undersampled k-space data by dictionary learning,'' \emph{IEEE Transactions
  on Medical Imaging}, vol.~30, no.~5, pp. 1028--1041, 2011.

\bibitem{2015_blind_CS_Ravishankar}
------, ``Efficient blind compressed sensing using sparsifying transforms with
  convergence guarantees and application to magnetic resonance imaging,''
  \emph{SIAM Journal on Imaging Sciences}, vol.~8, no.~4, pp. 2519--2557, 2015.

\bibitem{2016_Data_driven_Ravishankar}
------, ``Data-driven learning of a union of sparsifying transforms model for
  blind compressed sensing,'' \emph{IEEE Transactions on Computational
  Imaging}, vol.~2, no.~3, pp. 294--309, 2016.

\bibitem{2014_PANO}
X.~Qu, Y.~Hou, F.~Lam, D.~Guo, J.~Zhong, and Z.~Chen, ``Magnetic resonance
  image reconstruction from undersampled measurements using a patch-based
  nonlocal operator,'' \emph{Medical Image Analysis}, vol.~18, no.~6, pp.
  843--856, 2014.

\bibitem{2013_MRI_xiaobo}
B.~Ning, X.~Qu, D.~Guo, C.~Hu, and Z.~Chen, ``Magnetic resonance image
  reconstruction using trained geometric directions in 2{D} redundant wavelets
  domain and non-convex optimization,'' \emph{Magnetic Resonance Imaging},
  vol.~31, no.~9, pp. 1611--1622, 2013.

\bibitem{2014_TBME_Wang}
Y.~Wang and L.~Ying, ``Compressed sensing dynamic cardiac cine {MRI} using
  learned spatiotemporal dictionary,'' \emph{IEEE Transactions on Biomedical
  Engineering}, vol.~61, no.~4, pp. 1109--1120, 2014.

\bibitem{2007_PSF_Liang}
Z.~Liang, ``Spatiotemporal imaging with partially separable funcrions,'' in
  \emph{2007 4th IEEE International Symposium on Biomedical Imaging: From Nano
  to Macro}, 2007, Conference Proceedings, pp. 988--991.

\bibitem{2012_TMI_Zhao}
B.~Zhao, J.~P. Haldar, A.~G. Christodoulou, and Z.~Liang, ``Image
  reconstruction from highly undersampled (k, t)-space data with joint partial
  separability and sparsity constraints,'' \emph{IEEE Transactions on Medical
  Imaging}, vol.~31, no.~9, pp. 1809--1820, 2012.

\bibitem{2015_LR_sparse_MRI}
R.~Otazo, E.~Candes, and D.~K. Sodickson, ``Low-rank plus sparse matrix
  decomposition for accelerated dynamic {MRI} with separation of background and
  dynamic components,'' \emph{Magnetic Resonance in Medicine}, vol.~73, no.~3,
  pp. 1125--36, 2015.

\bibitem{2014_PLoSOne_Yu}
Y.~Yu, J.~Jin, F.~Liu, and S.~Crozier, ``Multidimensional compressed sensing
  {MRI} using tensor decomposition-based sparsifying transform,'' \emph{PLoS
  One}, vol.~9, no.~6, p. e98441, 2014.

\bibitem{2016_high_dim_MR_tensor}
J.~He, Q.~Liu, A.~G. Christodoulou, C.~Ma, F.~Lam, and Z.~Liang, ``Accelerated
  high-dimensional {MR} imaging with sparse sampling using low-rank tensors,''
  \emph{IEEE Transactions on Medical Imaging}, vol.~35, no.~9, pp. 2119--2129,
  2016.

\bibitem{2015_MIA_Zhang}
X.~Zhang, Z.~Xu, N.~Jia, W.~Yang, Q.~Feng, W.~Chen, and Y.~Feng, ``Denoising of
  3{D} magnetic resonance images by using higher-order singular value
  decomposition,'' \emph{Medical Image Analysis}, vol.~19, no.~1, pp. 75--86,
  2015.

\bibitem{2018_MIA_LRTV_algorithm}
J.~Yao, Z.~Xu, X.~Huang, and J.~Huang, ``An efficient algorithm for dynamic
  {MRI} using low-rank and total variation regularizations,'' \emph{Medical
  Image Analysis}, vol.~44, pp. 14--27, 2018.

\bibitem{1989_polynomial_Liang}
Z.-P. Liang, E.~M. Haacke, and C.~W. Thomas, ``High-resolution inversion of
  finite fourier transform data through a localised polynomial approximation,''
  \emph{Inverse Problems}, vol.~5, no.~5, p. 831, 1989.

\bibitem{2014_LORAKS}
J.~P. Haldar, ``Low-rank modeling of local \textit{k}-space neighborhoods
  ({LORAKS}) for constrained {MRI},'' \emph{IEEE Transaction on Medical
  Imaging}, vol.~33, no.~3, pp. 668--81, 2014.

\bibitem{2016_P_LORAKS}
J.~P. Haldar and J.~Zhuo, ``{P-LORAKS}: {L}ow rank modeling of local
  \textit{k}-space,'' \emph{Magnetic Resonance in Medicine}, vol.~75, no.~4,
  pp. 1499--1514, 2016.

\bibitem{2016_Off_the_grid}
G.~Ongie and M.~Jacob, ``Off-the-grid recovery of piecewise constant images
  from few fourier samples,'' \emph{SIAM Journal on Imaging Sciences}, vol.~9,
  no.~3, pp. 1004--1041, 2016.

\bibitem{2016_ALOHA}
K.~Jin, D.~Lee, and J.~Ye, ``A general framework for compressed sensing and
  parallel {MRI} using annihilating filter based low-rank {H}ankel matrix,''
  \emph{IEEE Transactions on Computational Imaging}, vol.~2, no.~4, pp.
  480--495, 2016.

\bibitem{2015_AC_LORAKS}
J.~P. {Haldar}, ``Autocalibrated loraks for fast constrained mri
  reconstruction,'' in \emph{2015 IEEE 12th International Symposium on
  Biomedical Imaging (ISBI)}, 2015, pp. 910--913.

\bibitem{2015_W_LORAKS}
------, ``Low-rank modeling of local k-space neighborhoods: from phase and
  support constraints to structured sparsity,'' in \emph{Wavelets and Sparsity
  XVI, International Society for Optics and Photonics}, vol. 9597, 2015, p.
  959710.

\bibitem{2018_Hu_LR_TMI}
Y.~{Hu}, X.~{Liu}, and M.~{Jacob}, ``A generalized structured low-rank matrix
  completion algorithm for mr image recovery,'' \emph{IEEE Transactions on
  Medical Imaging}, pp. 1--1, 2018.

\bibitem{1992_matrixPencil_Hua}
Y.~Hua, ``Estimating two-dimensional frequencies by matrix enhancement and
  matrix pencil,'' \emph{IEEE Transactions on Signal Processing}, vol.~40,
  no.~9, pp. 2267--2280, 1992.

\bibitem{2014_LR_Yuejie}
Y.~Chen and Y.~Chi, ``Robust spectral compressed sensing via structured matrix
  completion,'' \emph{IEEE Transactions on Information Theory}, vol.~60,
  no.~10, pp. 6576--6601, 2014.

\bibitem{2018_TBME_Hengfa}
H.~Lu, X.~Zhang, T.~Qiu, J.~Yang, J.~Ying, D.~Guo, Z.~Chen, and X.~Qu, ``Low
  rank enhanced matrix recovery of hybrid time and frequency data in fast
  magnetic resonance spectroscopy,'' \emph{IEEE Transactions on Biomedical
  Engineering}, vol.~65, no.~4, pp. 809--820, 2018.

\bibitem{2017_ACCESS_Guo}
D.~Guo, H.~Lu, and X.~Qu, ``A fast low rank hankel matrix factorization
  reconstruction method for non-uniformly sampled magnetic resonance
  spectroscopy,'' \emph{IEEE Access}, vol.~5, pp. 16\,033--16\,039, 2017.

\bibitem{2004_matrix_factorizations}
N.~Srebro, ``Learning with matrix factorizations,'' Thesis, 2004.

\bibitem{2010_SVT}
J.~Cai, E.~Candès, and Z.~Shen, ``A singular value thresholding algorithm for
  matrix completion,'' \emph{SIAM Journal on Optimization}, vol.~20, no.~4, pp.
  1956--1982, 2010.

\bibitem{2011_ADMM}
S.~Boyd, N.~Parikh, E.~Chu, B.~Peleato, and J.~Eckstein, ``Distributed
  optimization and statistical learning via the alternating direction method of
  multipliers,'' \emph{Foundations and Trends® in Machine Learning}, vol.~3,
  no.~1, pp. 1--122, 2011.

\bibitem{code_SPIRiT}
\BIBentryALTinterwordspacing
M.~Lustig and J.~Pauly, ``{SPIRiT}: {I}terative self‐consistent parallel
  imaging reconstruction from arbitrary k‐space,'' 2010. [Online]. Available:
  \url{http://people.eecs.berkeley.edu/~mlustig/software/SPIRiT_v0.3.tar.gz}
\BIBentrySTDinterwordspacing

\bibitem{code_ALOHA}
\BIBentryALTinterwordspacing
K.~H. Jin, D.~Lee, and J.~C. Ye, ``A general framework for compressed sensing
  and parallel {MRI} using annihilating filter based low-rank hankel matrix,''
  2017. [Online]. Available:
  \url{https://bispl.weebly.com/aloha-for-mr-recon.html}
\BIBentrySTDinterwordspacing

\bibitem{2004_TIP_Zhou}
W.~Zhou, A.~C. Bovik, H.~R. Sheikh, and E.~P. Simoncelli, ``Image quality
  assessment: {F}rom error visibility to structural similarity,'' \emph{IEEE
  Transactions on Image Processing}, vol.~13, no.~4, pp. 600--612, 2004.

\bibitem{2013_ESPIRIT}
D.~Bahri, M.~Uecker, and M.~Lustig, ``{ESPIRIT}-based coil compression for
  cartesian sampling,'' in \emph{International Society for Magnetic Resonance
  in Medicine}, 2013, Conference Proceedings, p. 2657.

\bibitem{2018_Vandermonde_Jiaxi}
J.~Ying, J.~Cai, D.~Guo, G.~Tang, Z.~Chen, and X.~Qu, ``Vandermonde
  factorization of {H}ankel matrix for complex exponential signal
  recovery—{A}pplication in fast {NMR} spectroscopy,'' \emph{IEEE
  Transactions on Signal Processing}, vol.~66, no.~21, pp. 5520--5533, 2018.

\bibitem{2017_TSP_Jiaxi}
J.~Ying, H.~Lu, Q.~Wei, J.~Cai, D.~Guo, J.~Wu, Z.~Chen, and X.~Qu, ``Hankel
  matrix nuclear norm regularized tensor completion for
  {\textit{n}}-dimensional exponential signals,'' \emph{IEEE Transactions on
  Signal Processing}, vol.~65, no.~14, pp. 3702--3717, 2017.

\end{thebibliography}

\end{document}